\documentclass[11pt]{article}

\makeatletter
\def\ps@top{\let\@mkboth\@gobbletwo
     \def\@oddhead{\rm\hfil\thepage\hfil}\def\@oddfoot{}
     \def\@evenhead{}\let\@evenfoot\@oddfoot}
\def\@bibsetup{\itemindent=-\leftmargin}
\def\@citesep{; }
\def\@cite#1#2{({#1\if@tempswa , #2\fi})}
\def\@biblabel#1{\hfill}
\def\thebibliography#1{\section*{References\markboth
 {REFERENCES}{REFERENCES}}\list
 {[\arabic{enumi}]}{\settowidth\labelwidth{[#1]}\leftmargin\labelwidth
 \advance\leftmargin\labelsep
 \usecounter{enumi}\@bibsetup}
 \def\newblock{\hskip .11em plus .33em minus -.07em}
 \sloppy
 \sfcode`\.=1000\relax}
\renewcommand{\section}{\@startsection {section}{1}{\z@}{-3.5ex plus -1ex minus 
    -.2ex}{2.3ex plus .2ex}{\centering\large\bf}}
\renewcommand{\subsection}{\@startsection{subsection}{2}{\z@}{-3.25ex plus
    -1ex minus -.2ex}{1.5ex plus .2ex}{\centering\bf}}
\makeatother
\begin{document}

\begin{center}
A McDonald Observatory Study of Comet 19P/Borrelly: \\
Placing the Deep Space~1 Observations into a Broader Context \\ [25pt]
Tony L. Farnham$^\dagger$ and Anita L. Cochran \\
University of Texas at Austin \\ Austin, TX 78712 \\ [20pt]
Accepted for publication in {\it Icarus} \\
\end{center}

\begin{center}
\vspace{1.0 in}
$\dagger$~ Current address: \parbox[t]{3in}{Department of Astronomy\\ University of
Maryland \\ College Park, MD 20742 \\ farnham@astro.umd.edu}
\vspace{0.5in}
\end{center}

\begin{center}{Abstract}\end{center}

We present imaging and spectroscopic data on comet 19P/Borrelly that
were obtained around the time of the Deep Space~1 encounter and in
subsequent months.  In the four months after perihelion, the comet
showed a strong primary (sunward) jet that is aligned with the
nucleus' spin axis.  A weaker secondary jet on the opposite hemisphere
appeared to became active around the end of 2001, when the primary jet
was shutting down.  We investigated the gas and dust distributions in
the coma, which exhibited strong asymmetries in the
sunward/anti-sunward direction.  A comparison of the CN and C$_2$
distributions from 2001 and 1994 (during times when the viewing
geometry was almost identical) shows a remarkable similarity,
indicating that the comet's activity is essentially repeatable from
one apparition to the next.  We also measured the dust reflectivities
as a function of wavelength and position in the coma, and, though the
dust was very red overall, we again found variations with respect to
the solar direction.  We used the primary jet's appearance on several
dates to determine the orientation of the rotation pole to be
$\alpha=214^\circ$,~$\delta = -5^\circ$.  We compared this result to
published images from 1994 to conclude that the nucleus is near a
state of simple rotation. However, data from the 1911, 1918 and 1925
apparitions indicate that the pole might have shifted by 5-10$^\circ$
since the comet was discovered.  Using our pole position and the
published nongravitational acceleration terms, we computed a mass of
the nucleus of $3.3\times10^{16}$~g and a bulk density of
0.49~g\,cm$^{-3}$ (with a range of $0.29 < \rho < 0.83$~g\,cm$^{-3}$).
This result is the least model-dependent comet density known to date.

\vspace*{10pt}

\newpage
\section{Introduction}

On 22~September~2001~UT, NASA's Deep Space~1 (DS1) spacecraft made a
close flyby of the nucleus of comet 19P/Borrelly, obtaining
high-resolution images, infrared spectra and particles and fields
measurements within about 12 hours of closest approach (Soderblom
{\it et al.}~2002).  The images that were obtained offer an unprecedented
look at the nucleus of this comet and promise to reveal many details
about the innermost region of the coma as well as the topology and
albedo of the nucleus.  However, due to the rapid velocity and short
duration of the encounter, additional information is needed to provide
a more global interpretation of the spacecraft measurements and how
they relate to observations of the entire coma.

In support of this mission, we utilized the 2.1-m and 2.7-m telescopes
of McDonald Observatory to observe comet Borrelly around the time of
the DS1 encounter and in subsequent months.  We obtained both
moderate-resolution spectroscopy and broad band imaging in R and V
filters.  In this paper we describe our observations and discuss how
they were used to analyze the morphology of the coma, probe the
asymmetric distributions of the gas and dust, and derive the
reflectivity of the dust.  We then highlight some unique inherent
physical characteristics of the comet and discuss how they were used
to determine the orientation of the spin axis.  Finally, we present an
analysis in which we used our pole determination and the comet's
nongravitational forces to constrain the mass and density of the
nucleus.  We also compare our results to those found from earlier
apparitions (Cochran and Barker 1999) and to other observers' results
from this apparition, including the DS1 measurements.

\section{Observations and Reductions}

We obtained two different types of data on comet Borrelly during the
2001 apparition: moderate-resolution spectroscopy and broad-band
imaging.  Table~\ref{log} is a log of our observations and a summary
of the geometric conditions.

\subsection{Spectroscopy}
The spectroscopic observations were obtained using the McDonald
Observatory Electronic Spectrograph No. 2 (ES2) on the 2.1-m Otto
Struve telescope.  The detector was a TI $800\times800$\,pixel CCD
with 15\,$\mu$m pixels that were binned by a factor of 2 in the
spatial direction.  The spectrograph is a long slit instrument with a
slit length of almost 200\,arcsec; each binned pixel subtends
1.89\,arcsec on the sky.  For the observations of the comet and solar
analogue stars, the slit was 2.1\,arcsec wide.  The slit was widened
to 12\,arcsec for spectrophotometric standard stars in order to ensure
no loss of light.  The nominal 2-pixel resolution was 6.5\AA.  The
spectra covered the bandpass from 3043--5680\AA\ in September and
3220--5872\AA\ in November.

The reduction of the data proceeded in a standard manner with removal
of the bias using a masked region of the chip and of the
pixel-to-pixel sensitivity variation using observations of an
incandescent (flat field) lamp.  Wavelength calibrations were
performed with spectra of an argon lamp observed at the same position
as the objects.  Flux calibration was accomplished with observations
of spectrophotometric standards (Stone 1977). 

In normal operation, the slit is oriented east/west.  However, this
orientation can be altered by rotating the instrument on the
telescope.  The position angle (PA) of the slit on the sky for the
observations is noted in Table~\ref{log}.  Guiding was accomplished by
imaging the slit with a CCD guide camera so that we were guiding
directly on the cometary image.  Care was taken to observe the argon
lamp at each slit orientation to map any possible motions of the
spectrum on the detector produced by rotating the instrument.

The spectrum of any comet consists of a combination of molecular
emissions, generally from resonance fluorescence, superimposed on a
solar spectrum reflected from the dust.  In order to study the
distribution of the gas in the coma, we removed the underlying
continuum and then integrated across the emission band wavelengths.
We used observations of solar analogue stars obtained during the same
observing runs to represent the solar spectrum.  Complete details of
our analysis procedures can be found in Cochran {\it et al.} (1992).
The bandpasses for the gas and continuum can be
found in Table~I of that paper and the constants for conversion of the
band intensities to column densities can be found in Table~II of that
paper.

Since the ``apertures'' of a slit spectrograph are very small, with
most of the apertures not including the optocenter, the standard
$Af\rho$ formalism of A'Hearn {\it et al.} (1984) is not very useful
for studying the dust with our spectra.  These problems were discussed
in detail by Storrs {\it et al.} (1992).  
Instead of using the $Af\rho$ formalism, we studied the dust in the
coma by determining the flux in the continuum regions of the spectrum
without removing the solar spectrum.  Some of these bandpasses are
slightly contaminated by gas but the complex nature of the cometary
spectrum makes observing true continuum difficult.  However, with the
use of a spectrograph, we can isolate continuum regions more easily
than with narrow-band filters.  We can also ratio the derived fluxes
to the observed solar flux (from observations of the solar analogue)
in the same bandpasses in order to determine the relative reflectivity
of the dust.

\subsection{Imaging}

We obtained imaging data on a total of five observing runs at McDonald
Observatory, under the geometric conditions listed in Table~I.  The
December~2001 data were obtained with the 2.1-m Otto Struve Telescope,
and data from the other four runs were obtained with the 2.7-m Harlan
J. Smith telescope.  In both cases, the Imaging Grism Instrument, a
5:1 focal reducer, and a TeK 1024$\times$1024 CCD (which is partly
vignetted on both telescopes) were used.  On the 2.7-m, this
configuration produces a 7\,arcmin field with 0.57\,arcsec pixels.  On
the 2.1-m, the result is a 6\,arcmin field with 0.48\,arcsec pixels.
During the September and November runs and on 6-7~December, the seeing
was typically between 1 and 2\,arcsec (FWHM); on 4~and 5~December, it
was highly variable, ranging from 2.5 to 4\,arcsec; during the
February run, it was typically around 2.5\,arcsec; and in May it was
about 3\,arcsec.

Processing of the CCD images followed standard procedures and was done
using the CCD reduction packages in the Image Reduction and Analysis
Facility (IRAF).  The bias was removed in two steps, first applying
the overscan region to remove the bulk value, and then subtracting a
master bias frame, created by averaging many bias images, to remove
the residual for each individual pixel.  Flat fielding was done using
twilight sky flats that were medianed together to remove stars.

Many of the images were obtained under non-photometric conditions, but
for all except the May observing run, at least one night per run was of
sufficient quality to provide an absolute calibration.  On these
nights, Landolt standards (Landolt~1992) were observed at several
airmasses and in both filters so the extinction per airmass and color
coefficients of the extinction could be determined, as well as the
zero-point offset of the instrumental magnitudes.  Using the standard
star information, the comet images were converted to the standard
magnitude system, and ultimately to absolute fluxes.  Finally, in
order to compare the inherent brightness levels of the comet between
the four observing runs, the fluxes were converted to total
luminosities by multiplying by $4\pi\Delta^2$.  No calibration was
done for the May data.

The December observing run presented a couple of problems that
affected the quality of the data.  The 2.1-m telescope does not track
very well at cometary rates, and even though short exposures were used
to avoid significant trailing, there are still guiding problems in
some of the images.  (Note that the tracking problems did not affect
the spectra, because, in those observations, we guided actively on the
comet image on the slit.)  The two photometric nights from the
December run had poor seeing, while the two nights that had decent
seeing were non-photometric.  To produce the best representative
images from this run, we calibrated the images from 4~and 5~December
using the standard star measurements (being careful to account for
seeing variations during each night) and then used the total
brightness of the coma to calibrate the images from 6~December.  This
makes the inherent assumption that the comet's brightness is not
changing on daily timescales, but a comparison of the images from
4~and 5~December supports this assumption.  It is possible that a
minor outburst could have occurred between December~5 and~6, but there
is no morphological evidence for this.  

\section{Results}

\subsection{Appearance and Evolution of the Coma}\label{sec:appear}

The appearance of comet Borrelly, as shown in representative images in
Figure~\ref{images}, provides a record of the temporal evolution of
the coma during the months after perihelion.  In September (during the
DS1 encounter) the inner coma was dominated by a straight, narrow
feature that pointed to within 7$^\circ$ of the sunward direction as
projected onto the plane of the sky.  We know that this feature is not
a classical anti-tail, which is produced by projection effects of the
dust tail as the Earth passes through the comet's orbit, because the
Earth was 19.5$^\circ$ above the comet's orbital plane at the time.
Therefore, it must be a jet, produced by an isolated active region on
the comet's surface, and we have designated it the ``primary jet.''
The width of this jet was about 35$^\circ$ (FWHM).  A dust tail was
obvious in the anti-solar direction but it was noticeably fainter than
the jet.  Throughout the following months, the general morphology was
similar, but the jet seemed to fade relative to the rest of the coma.
In November, the primary jet was closely aligned with the sunward
direction, but its brightness had dropped to the point that it was
comparable to that of the dust tail.  By December, the jet was fainter
than the dust tail and was pointed at an angle about 15$^\circ$ from
the apparent sunward direction.

Measurements by Lamy {\it et al.} (1998) and Mueller and Samarasinha
(2002)  
indicate that the comet is rotating with a
period of 25--26\,hours.  Given the time span of our observations in
September and December, we have coverage of about one-quarter of a
rotation on each run, so we should expect to see evidence of
rotation in our images.  However, we see no indication that the coma
was changing, on timescales of either hours or days.  We searched for
evidence of temporal variations (e.g., a change in the jet's PA or a
non-radial morphology formed by the material spiraling outward from
the rotating nucleus) but found none.  Even with the application of
image enhancement techniques, including 1/$\rho$ removal, digital
filters and unsharp masking, we could see no apparent change in the
jet PA or deviations from a purely radial morphology.  It is possible
that the rotation axis could be oriented such that, in one of the
observing runs, the jet might show little change (for example, if the
axis were parallel to the plane of the sky and the jet were rotating
along the line of sight).  However, changes in the viewing
geometry from September to December make it extremely unlikely that
this could occur during all three of these observing runs.  Therefore,
the lack of any movement or curvature in the primary jet leads us to
conclude that it must be situated close to the rotation pole of the
nucleus.  Samarasinha and Mueller (2002)  
come to this
same conclusion and invoke dynamical arguments to further conclude
that the nucleus must be very close to a state of principle axis
rotation.  We address the comet's rotation state in more detail in
Section~\ref{section:pole}.

By February, the comet's appearance had changed somewhat.  A
well-defined, curved jet was visible, extending out from the nucleus
at a position angle of $\sim335^\circ$, roughly perpendicular to the
projected sunward direction.  Is this jet simply a different
projection of the primary jet, or is it a new active region not
previously seen?  We can address this question using the basic
geometry of the comet's orbit.  The inset panels in
Figure~\ref{images} show an inertial reference frame based on the
comet's orbital plane (the X axis points in the anti-solar direction
at perihelion and the Z axis is parallel to the orbital angular
momentum vector).  In the September, November and December images, the
primary jet consistently extended in the --X direction, while the jet
observed in February clearly lies in the +X direction.  Because the
projection effects shifted only gradually during our observations,
with remarkably little change from December through May, we conclude
that the feature seen in February is a secondary jet that points in
the opposite direction from the primary jet.  The changing relative
brightnesses of the two jets suggest that seasonal effects caused the
primary active region to shut down while the secondary became more
active.  The curvature of the secondary jet can be attributed either
to the spiral effect produced by rotation of the nucleus or to the
effects of radiation pressure, which is acting roughly perpendicular
to the jet axis.  Unfortunately, our February observations were not
extensive enough to look for motion of the jet as a function of time,
so we cannot differentiate between these possibilities.  

The February image also shows a protrusion of the coma in the
southeast direction.  At first glance, this extension appears to have
the same radial structure as the primary jet in the September,
November and December images, suggesting that the primary jet might
continue to be active.  Upon further inspection, however, it is clear
that the February feature is not truly radial, but appears to be a
linear structure that is offset from the nucleus.  To illustrate this,
a contour plot is shown in Fig.~\ref{images}, with a dashed line
connecting the points of the protrusion for each contour level.  At
large cometocentric distances, the protrusion is nearly linear, but is
offset so that if it were extended, it would not intersect the
nucleus.  Only at smaller distances, where the central coma dominates
the brightness, does the dashed line curve in toward the nucleus.  The
fact that the protrusion does not extend radially in to the nucleus
means that it is not produced by continuous activity from the primary
jet in the same manner as it was from September through December.
Instead, the protrusion must be composed of relatively large dust
particles that were emitted from the primary jet earlier in the
apparition and are only slowly being blown away from the sun by
radiation pressure.  Schleicher {\it et al.}~(2002) observed this same
protrusion in their January data and came to the same conclusion.

Finally, Fig.~\ref{images} also shows a picture from the May observing
run.  Even though the image quality is poor, the secondary jet can be
seen at a PA of 308$^\circ$.  Again, our data set is too limited to
search for changes in the jet as a function of time.  Also, the
signal-to-noise is too low to unambiguously follow the jet to more
than about $2\times10^4$~km from the nucleus, which means we can't
look for spiral structures that would provide clues to the rotation
period or the location of the active region.

\subsection{The Distribution of the Gas in the Coma}

Our spectroscopic bandpass allows for the detection of CN, C$_{2}$,
C$_{3}$, CH and NH and also for detection of OH in September and 
NH$_{2}$ in November.  For each spatial pixel
(1.89\,arcsec) along the extent of the slit, we can measure the band
intensities of the molecules and convert them to column densities. 
Figure~\ref{imageslit} shows images of the comet with the positions of the
slit marked on them for the observing runs for which we obtained spectra.
We can then use the spectroscopic data to probe the distribution of the
gas along certain vectors.  Figure~\ref{spectrum} shows the spectrum
of comet Borrelly at the optocenter location and 20,000\,km north of
the optocenter.  For comparison, another cometary spectrum obtained
with a comparable instrument is shown.  Note that the Borrelly
optocenter spectrum shows more continuum than the off-optocenter
spectrum.  Compared with comet C/1996~B1 (Szczepanski), the
C$_{2}$ and C$_{3}$ in Borrelly are less strong relative to the CN.

Figure~\ref{gas} shows the derived column densities as a function of
cometocentric distance for C$_{2}$ and CN on each night that spectra
were obtained.  At least two spectral images were obtained on each
night, with five images obtained on 22 September.  The column
densities from each spectral image are denoted with a different
symbol.  The optocenter of the comet was always imaged on part of the
slit, and the colors in the plot indicate on which side of the
optocenter a column density was measured.

Several points are noteworthy from this figure.  There is excellent
agreement among the column densities derived from the different
spectral images on each night.  Indeed, one can estimate a low level
of uncertainty in the measured column densities from the small spread
in the data.  When the slit was oriented along the the Sun/anti-Sun
line (parallel direction), the gas shows a highly asymmetric
distribution, which is much larger than the uncertainties in the data.
However, when the slit was aligned perpendicular to this line
(perpendicular direction), the gas distribution seems quite symmetric.

We have fit a Haser model (Haser 1957) to the data from 23 September
and the distribution perpendicular to the Sun/anti-Sun direction on 21
November in order to derive the production rates, $Q$.  Although these
models assume spherical symmetry, which is clearly \underline{not} in
evidence for Borrelly, we are simply using the Haser fits as a
comparison tool for visualizing the differences in the different
distributions.  We adopted the Haser model scalelengths of Randall
{\it et al.} (1992), which were also adopted by A'Hearn {\it et al.}
(1995).  These scalelengths are different from those we used
previously (Cochran and Barker 1999), but they fit the September data
better than the previous values.  We adopted a constant velocity of
1\,km\,sec$^{-1}$, so we have actually derived $Q/v$.  The values of
$Q/v$ were chosen for the best fit to all of the data, weighted by the
signal/noise of the column densities.  The derived Haser production
rates are indicated in the appropriate panels of the plot and are
listed in Table~\ref{prodrates}.

The Haser fit to the CN data of 23 September is excellent.  For
C$_{2}$ on 23 September, the Haser model overpredicts the column
densities in the inner and outer coma and underestimates the column
densities at around 10,000\,km.  This slight mismatch is the result of
the simplicity of the Haser model, which assumes a simple
parent-daughter process; C$_{2}$ is probably a granddaughter species.
The Haser fit to the CN data of 21 November is slightly worse than for
23 September but is still acceptable.  However, the C$_{2}$ data of 21
November are fit poorly.  

The Haser fits for 23 September are transposed to the data of 22 and
25 September in order to guide the eye for a comparison of the gas
distribution from night to night.  Similarly, fits from the 21
November perpendicular slit profiles are plotted on the 19 November
and 21 November parallel slit data.  There is no rescaling of the
production rates.  The distribution of the gas on 25 September is
comparable to that on 23 September, and both of these nights have a CN
gas distribution that is intermediate to the sunward and anti-sunward
distributions of 22 September.  The anti-sunward C$_{2}$ distribution
looks very much like the perpendicular C$_{2}$ gas distribution, but
the fit does not model well the sunward C$_{2}$ distribution.  Note in
particular the C$_{2}$ column densities on the sunward side on 21
November.  This gas distribution is essentially flat out to 50,000\,km
from the optocenter.  This is a very unusual distribution and it is
obvious that no simple model can fit these data.

Figure~\ref{compare} is a comparison of our C$_{2}$ and CN
measurements from September 2001 with those obtained by Cochran and
Barker (1999) in November 1994.  The 1994 data were obtained with the
same detector but with a different spectrograph on the 2.7-m
telescope. They cover a similar bandpass at only slightly lower
resolving power.  The comet was at almost exactly the same
heliocentric distance during these two data sets (1.36~{\sc AU} in
1994, 1.37~{\sc AU} in 2001), though the geocentric distance differed
significantly (0.7\,{\sc au} and 1.5\,{\sc au}).  Fortuitously, with
the exception of the geocentric distance, all of the geometric
conditions were virtually identical for these two sets of data,
including the phase angle (41.2$^\circ$ and 43.6$^\circ$), the orbital
latitude of the Earth (19.5$^\circ$ and 20.3$^\circ$) and even the
right ascension and declination!  Next to the dates of observation in
Figure~\ref{compare}, we note the corresponding number of days
relative to perihelion.  As with the September 2001 data, the November
1994 data were obtained with the slit both parallel and perpendicular
to the Sun/anti-Sun line.  We wish to emphasize that no rescaling was
done in plotting the data from the two apparitions in this figure, yet
the agreement between the two sets is remarkable, especially for the
data obtained parallel to the Sun/anti-Sun line.  Indeed, one could
say that there were no differences between these two apparitions.  For
the data obtained perpendicular to the Sun/anti-Sun line, the column
densities in the inner coma agree quite well, but the 2001 data seem
to show a steeper decline at larger distances.  This trend is seen in
both the CN and C$_2$ profiles.  We know that the deviation is not the
result of incorrect instrumental plate scales because of the excellent
match of the distributions along the Sun/anti-Sun direction. In
addition, a scale error of 30\% would be needed to make the
perpendicular data match at 10$^5$\,km from the optocenter, and this
large an error is not plausible.

Why does the density in 2001 fall off more quickly than in 1994?  The
excellent agreement in the inner coma indicates that the production
rates at the nucleus are quite consistent, so the differences must be
the result of processes occurring in the coma.  During 2001, the Sun
was just past solar maximum, while during 1994 the Sun was relatively
quiescent.  A more active Sun causes the lifetimes of the
photodissociation products to decrease and thus shortens the
scalelengths.  This is consistent with what we observed and suggests
that differences in solar activity may be responsible for the
differences seen in these measurements.

Since the C$_{3}$ is generally weaker and spread over a wider bandpass
than the CN or C$_{2}$, C$_{3}$ column density distributions generally
show more scatter than those for C$_{2}$ and CN.  This is especially
true of our Borrelly spectra because the signal from C$_{3}$ is quite
weak.  This weakness leads to much more scatter in the data and it is
impossible to tell whether or not the C$_{3}$ gas distribution is
symmetric along the Sun/anti-Sun line.  The C$_{3}$ production
rates are included in Table~\ref{prodrates}.  Though our
wavelength coverage also includes emissions of other molecular
species, our column densities have sufficiently low signal/noise that
we could not say anything meaningful about the gas distribution for
these species, nor could we derive production rates.

Inspection of the values in Table~\ref{prodrates} shows that comet
Borrelly is mildly depleted according to the definitions of A'Hearn
{\it et al.} (1995).  This is in accord with A'Hearn {\it et al.}'s
findings for Borrelly, as well as with the results of Cochran and
Barker (1999) and Schleicher {\it et al.}~(2002).

\subsection{The Distribution of the Dust}\label{sec:dust}

Our gas observations show a clear asymmetry in the Sun/anti-Sun
direction that is not seen in the perpendicular direction.  Because
the gas carries dust particles off the nucleus, we expect to observe
similar characteristics between the gas and dust distributions.
Figure~\ref{dustflux} shows the measurements of the average flux
within each continuum bandpass from the September spectra.  We did not
measure the continuum in the last spectral image from each night
because the sky was beginning to brighten during these exposures, and
the continuum is contaminated by the sky flux.  The 4150--4175\AA\
bandpass suffers from contamination from C$_{3}$; however, the C$_{3}$
band is weak and is important only near the nucleus.  The other
bandpasses contain little contamination, though the data from the
3715--3770\AA\ bandpass have a great deal of scatter because the
signal from the continuum is quite weak at this wavelength.

Inspection of Figure~\ref{dustflux} shows that the asymmetry seen in
the Sun/anti-Sun gas observations is also present at all wavelengths
of the continuum measurements, even the noisy 3715--3770\AA\ band.  In
the perpendicular direction, there are fewer spectra, and though it
appears that some asymmetry may also be present, it is less certain.
The solid lines on these plots represent a $\rho^{-1}$ falloff, where
$\rho$ is the cometocentric distance projected on the sky.  In the
perpendicular direction at all wavelengths (with the possible
exception of the bluest band, which is noisy), the dust declines more
steeply than $\rho^{-1}$.  Along the sunward (jet) direction, the dust
follows a $\rho^{-1}$ dependence out to 30,000\,km.  Beyond
this distance, it appears to drop more steeply, but the noise also
increases.  The anti-sunward direction shows a $\rho^{-1}$ decline or
slightly steeper.

In the imaging data, the V and R bandpasses contain flux from both
continuum and gas, with the contamination of the gas to the V filter
flux being somewhat greater than in the R filter.  However, with the
low levels of gas in this comet, the continuum should dominate the
surface brightness of images obtained with these filters.  We made the
assumption that, to first order, all of the surface brightness in our
images is produced by dust, and we used representative images to derive
radial profiles in different directions.  Comparing the radial
profiles obtained from the images to those measured in the spectral
continuum regions shows a good match, which indicates that the images
are indeed dominated by the dust.  In addition, the profiles derived
for the V and R images look almost identical, which also indicates
that dust dominates the surface brightness.  Because of the broad
filter bandpass, the signal/noise of the images is higher than in the
flux measures from the spectra; therefore, we can use the profiles
from the images to examine in more detail the structure of the coma in
the ``linear'' portion of the distribution from 3,000--30,000\,km.

Jewitt and Meech (1987) studied the comae of 10 comets by measuring
the falloff in the surface brightness as a function of projected
cometocentric distance (which they quantified by assuming a simple
relation $B\propto\rho^{m}$).  As part of this analysis, they
generated Monte Carlo dust models to show that a steady-state outflow
of dust from the nucleus would produce a canonical radial distribution
that declines as $\rho^{-1}$.  Alternatively, a slope that deviates
from $m=-1$ indicates that the dust outflow is being influenced by one
or more factors: radiation pressure acting on the dust, temporal
changes in the optical properties of the grains ({\it e.g.}, grains
are sublimating, fading or fragmenting) and/or temporal variations in
the emission from the nucleus.

In Figure~\ref{radprof4} we show radial distributions in four
directions for a representative image from each of the first four
observing runs.  (Due to the low S/N and lack of calibration, the
profile from the May data was not included here.)  We used the
dominant jet in each image (the primary jet for the first three runs,
the secondary jet in the February run) to define our reference
direction.  Note that the primary jet is a few degrees off sunward, so
the jet radial profiles are not identical in direction to profiles
from the spectra, which were aligned with the sunward direction.  For
each date, cuts were taken along the jet, in the direction opposite
the jet, and in the two perpendicular directions.  In the February
profile, the radial cut (the thin line in Figure~\ref{radprof4}) was
taken at a PA of 335$^\circ$, while the thick line represents the
profile that follows the approximate center of the curved jet.  The
resulting profiles on all dates are plotted at the same scale so the
luminosities, as well as the slopes of the radial distributions, can
be compared directly.  The discussion below focuses on the profiles in
the region outside of about 4,000~km, where the seeing and tracking
uncertainties have less effect, and inside of 20,000 km, where the
signal/noise is high.

If we initially focus on the first three observing runs, inspection of
the radial profiles shows some interesting results, with only subtle
changes with time.  First, the profile along the primary jet has a
slope of $m=-1$ in September and became only slightly shallower in
the next two observing runs, while the tail (anti-jet) direction
maintained a slope of $m=-1$ throughout all three runs.  The north and
south perpendicular profiles both exhibited slopes of $m=-1.3$ in
September and then became shallower ($m=-1.2$) in November.  In
December, however, the southern measurement retained its $m=-1.2$
slope, while the northern measurement steepened again to $m=-1.3$.

It is clear that the jet and the tail exhibited the canonical $m=-1$
falloff, while in the perpendicular directions something is affecting
the outflow.  We believe that radiation pressure, combined with the
near alignment of the jet with the sunward direction, can be used to
qualitatively explain the different radial profiles in these
observations.  The dust emitted from the jet is contained in a narrow
(35$^\circ$) cone that points in the general direction of the Sun (the
projected direction is within 15$^\circ$ of the sun in each case).
The dust that flows into this cone initially exhibits a $\rho^{-1}$
falloff, as is seen in the jet measurement.  Ultimately, radiation
pressure overcomes the outward momentum and the grains are turned
around and pushed back toward the nucleus.  Because the cone is narrow
and is pointed near the direction of the sun, much of the dust will
pass very close to the nucleus as it flows into the tail (at least as
seen from the Earth).  The rest of the dust will pass at larger
cometocentric distances, but the falloff will be relatively steep, as
is observed in the perpendicular directions.  We also cannot rule out
the possibility that there is some isotropic emission from the
nucleus.  If so, this material would provide an additional
contribution to the radial profiles.

In the February image, the distribution of the dust creates unusual
radial profiles.  In Figure~\ref{radprof4}, two separate measurements
are given for the secondary jet.  The first, shown with a thin line,
is the true radial profile extending outward from the nucleus at a PA
of 335$^\circ$.  At small projected distances, this profile is aligned
with the center of the jet and exhibits a $\rho^{-1}$ falloff.
However, at larger distances, the jet curves away from the sunward
direction, causing the straight radial profile to effectively move
away from the center of the jet and ultimately off of it altogether.
In this outer region, where the jet no longer contributes, the profile
drops rapidly with a slope $m<-1.3$.  The other profile from the
secondary jet, shown with a thick line, is not a true radial profile
with respect to the nucleus, but instead follows the curvature of the
center of the jet. In this case, the profile exhibits a $\rho^{-1}$
decline to beyond 30,000~km.  This suggests that the secondary jet is
constantly emitting dust with little temporal variations ({\it i.e.},
the illumination of the source does not change much on a scale of a
half day during this time) and the dust is being deflected by
radiation pressure.  The anti-jet profile and the South perpendicular
directions, which both lie along the broad dust tail to the South and
West, have profiles with slopes near $m=-1$.  This gradient reflects
the dispersion of the dust as it spreads out down the tail.  Finally,
the North perpendicular direction falls off in the same manner as the
(true) radial profile along the secondary jet, which indicates that we
are seeing a high-density region near the nucleus (from
isotropic emission?) that falls off very rapidly as solar radiation
pressure pushes the dust in the opposite direction.

Comparison of the luminosities in the four panels in
Figure~\ref{radprof4} shows that comet Borrelly faded rapidly during
the time of our observations.  By November, the brightness had dropped
by a factor of 4 from its September level.  Interestingly, the comet
remained at about this same brightness in December, but by
February, it had faded by another factor of 2.5.  This fading is much
too rapid to be explained by the $R_h^{-2}$ decline that is expected
as the comet moves away from the sun.  We are seeing either a
depletion of volatiles or reduced production rates due to seasonal
effects.  As discussed above, the production rates in 2001 are the
same as in 1994, so it is highly unlikely that volatiles would happen
to be depleted as we observed the comet on this apparition.
Therefore, we conclude that the rapid fading is due to seasonal
effects.  This is discussed further in section~\ref{section:pole}.

In general, we can make several comments regarding the radial profiles
from our images.  First, the relative brightnesses of the radial
profiles indicate that a large fraction of the material in the coma is
released from the two jets that are visible in our images (though not
necessarily from both jets at the same time).  Second, in order to
produce the observed $\rho^{-1}$ falloff that we see in all the jet
profiles, we know that the source regions must be in a steady state of
emission, which suggests that they are in direct sunlight for at least
most of a rotation during the times of our observations.  Finally,
radiation pressure is acting, sometimes severely, on the dust grains,
as is evident from the radial distributions that deviate from $m=-1$
and the curvature of the jet and the offset protrusion in the February
image.  Unfortunately, because most of the material is being emitted
from isolated active regions, detailed models of the dust motions will
be necessary to extract information about the particle size
distributions and emission velocities.

\subsection{Dust Colors}\label{sec:color}

Returning to the spectral observations, we can utilize the dust
continuum measurements, along with the observations of the solar
analogue stars (with the same instrumental setup used for the comet
observations) to determine the color of the dust in the inner coma.
By computing the flux in the same bandpasses for the stars as for the
comet, we can obtain the reflectivity of the dust.  (Even at the
optocenter the dust coma dominates over the nucleus contribution in
our spectra.)  The reflectivity is then found from the ratio of the
cometary and stellar continuum fluxes.  We normalize the reflectivity
to 1 at our reddest wavelength, centered at 5245\AA.

The mean optocenter reflectivities for September and November were
determined by averaging the derived reflectivity for the pixel
containing the optocenter in all of the spectral images from each run.
The optocenter pixel includes the light from the inner coma as well as
from the nucleus itself.  If we assume that we are observing the
broadest side of the nucleus and that it has a uniform albedo of 3\%,
the highest measured by DS1, we can estimate that the contribution of
the nucleus flux to the total flux in this pixel is of order 8--15\%.
Both assumptions imply that we are seeing the absolute maximum
possible contribution from the nucleus, which, given the rotation and
albedo variations, is not likely to be the case.  Therefore, this
estimate of the flux is an upper limit for what we can expect as the
contribution from the nucleus.

There were eight total optocenter reflectivities from September and
six from November.  The average reflectivities are shown in
Figure~\ref{reflectivity}.  For our wavelength range, the optocenter
reflectivities are quite red (almost as red as Pholus, though this is
a comparison of the dust in Borrelly to the surface of Pholus).  The
error bars are the standard deviation of the reflectivities of each
image from the mean and thus represent our formal error.  Though the
November and September optocenter colors are essentially the same,
within the error bars, we note that the November reflectivities are
consistently greater at all wavelengths than those in September,
suggesting that the optocenter color might have been slightly less red
in November than in September.

Also included in Figure~\ref{reflectivity} are off-optocenter
reflectivities derived for various orientations for the September
data.  Since the continuum declines rapidly with cometocentric
distance, we were not able to measure accurately the reflectivities
far from the optocenter.  Instead, we averaged the three pixels
adjacent to the optocenter in a particular orientation from all
spectral images containing that orientation.  Thus, each reflectivity
value for the off-optocenter data in this figure represents an average
of 12 reflectivities spanning a range from around 5,000--17,000\,km
from the optocenter.  The bluest bandpass was too noisy to be
meaningful.  The off-optocenter reflectivities are all even redder
than the optocenter reflectivities.  The two reflectivity curves from
the orientations perpendicular to the Sun/anti-Sun direction agree
quite well with one another; the tailward color is even redder than
the perpendicular directions; and the sunward direction is the least
red, though the sunward dust is slightly redder than the optocenter
dust.  It is unclear whether the turn-up at the 4150\AA\ bandpass is
real in these curves; C$_3$ contamination of the bandpass may be
contributing to the flux, producing an artificial enhancement.

Any aperture will contain a mixture of particle sizes, but the red
color indicates that the optically dominant particles must be slightly
larger than the wavelengths we are observing.  While the sunward dust
could be considered to be the same color as the optocenter dust to
within the error bars, the tailward dust and the dust perpendicular to
the Sun/anti-Sun line are clearly redder, and so represent scattering
from larger dust grains on average.  Radiation pressure effects should
push the smallest particles down the tail faster than the larger
particles, which would cause the tail to appear bluer.  This is
contrary to what is observed, suggesting the jet is producing
particles with a mass distribution that favors smaller particles than
are seen in the tail.  The fact that the optocenter is the same color
or bluer than the jet supports the idea that the jet is producing
small particles since the optocenter pixel should contain a higher
percentage of the jet than the off-optocenter pixels.  It would be
necessary to obtain reflectivities over larger cometocentric distances
and over a larger spectral range to quantify the size distribution.

\subsection{Pole Orientation Determination}\label{section:pole}

The high-resolution images obtained by the Deep Space~1 spacecraft
revealed that the nucleus was highly elongated, with dimensions of
$4\times4\times8$\,km (Soderblom {\it et al.} 2001).
The images also show a number of active regions
near the narrow waist of the nucleus, though a large, highly
collimated jet strongly dominates the emission.  The flyby occurred
too rapidly for the spacecraft data to place any constraints on the
rotation of the nucleus, but as mentioned earlier, Hubble Space
Telescope (Lamy {\it et al.}  1998) and ground-based observations
(Mueller and Samarasinha 2002) indicate that the comet's nucleus is
rotating with a period of 25--26\,hours.   
If we
make two basic assumptions, we can use our imaging data from
September, November and December to determine the orientation of the
spin axis of the nucleus, thus adding to the overall picture of the
nucleus properties.

First, we assume that the nucleus is near a state of simple rotation.
This assumption is supported by the dynamical arguments presented by
Samarasinha and Mueller (2002)  as well as by the
gas profiles from 1994 and 2001 (Figure~\ref{compare}).  The profiles
for the two apparitions are remarkably similar, even to the extent
that the sun-tail asymmetry matches.  This agreement, combined with
the fact that the data were obtained under nearly identical geometric
conditions, strongly suggests that the pole orientation was the same
for both apparitions.  If the nucleus had any significant complex
rotation or precession of its angular momentum vector, it would be
highly implausible that the spin axis would return to the same
orientation at the same time that all of the other geometric
conditions matched and we were observing the comet again.  Our second
assumption is that the primary jet is on or very near the rotation
pole, with the dust emission aligned with the spin axis.  We believe
this is a good assumption for the reasons discussed in
section~\ref{sec:appear}.  We also note that, for our purposes, the
jet, which is 30--35$^\circ$ wide, could be 5--10$^\circ$ from the
pole without affecting the solution.

Using the above information, we know that the position angle of the
center of the jet defines the projection of the spin axis onto the
plane of the sky.  In three dimensions, however, the pole can lie
anywhere along a plane defined by the jet PA and the line of sight
(LOS), so one set of observations cannot define the pole position
uniquely.  Incorporating data from a second observing run, where the
observing geometry has significantly changed, can resolve the
ambiguities.  Because of the different geometry, the pole will appear
to lie along a different plane, and the intersection of planes from
the different observing runs defines the actual pole orientation in
inertial space.  Additional observing runs can be incorporated to check for
consistency.

To measure the position angles of the jet, we first performed a plate
solution to measure any rotation of the image from a North-South
orientation.  Next, we processed the images with a 1/$\rho$
enhancement and then unwrapped the coma images from $X,Y$ format into
a $\rho,\theta$ format, where each line represents a constant
cometocentric distance and each column represents a constant PA.  With
this format, it was a simple matter to plot a line of data and measure
the value of $\theta$ at which the brightness peaked, giving a measure
of the central PA of the jet.  By measuring the PA
at different radial distances (different lines) we made one last check
for changes in the jet's position as a function of $\rho$ and saw no
indication that the jet was not aligned with the spin axis.  Our
results showed that the jet's PA was constant, to within the
uncertainties, out to a distance greater than 100~pixels
($\sim$50,000~km) on each date.  For the three observing runs in 2001,
we obtained jet PAs of: Sept, 93$^\circ\pm2^\circ$; Nov,
115$^\circ\pm2^\circ$; and Dec, 131$^\circ\pm5^\circ$.  The larger
uncertainties in the December data reflect the fact that the jet was
not as bright or well-defined as it was on earlier dates.

Figure~\ref{poleplot} shows the pole/LOS planes for the different
epochs projected onto the celestial sphere.  One curve is plotted per
day during each run, producing three nearly overlapping curves for
September, one for November and four for December.  The optimum
intersection point, weighted by the errors at each epoch, is
$\alpha=214^\circ,~\delta = -5^\circ$, with an uncertainty of
4$^\circ$ overall.  The agreement between the three epochs is
excellent, which indicates that the pole solution is consistent for
all three dates.  Because the jet is located at the pole, we have no
information about the sense of the rotation, so we have arbitrarily
defined the North pole to be the one aligned with the primary jet.
(Schleicher {\it et al.}~(2002) present evidence from the secondary
jet to support the fact that this is indeed the North pole, by the
right hand rule for rotation.)  Our pole solution is consistent with
the $\alpha=214^\circ,~\delta =-6^\circ$ position obtained by
Schleicher {\it et al.} and very close to the $\alpha=218.5^\circ$,
$\delta=-12.5^\circ$ position (uncertainty of 3$^\circ$ in each
direction) found from the DS1 images (Soderblom {\it et al.} 2002) and
the $\alpha=221^\circ,~\delta = -7^\circ$ position found by
Samarasinha and Mueller (2002) (which is not very well constrained in
one direction).

Figure~\ref{subearth} shows the sub-solar and sub-Earth latitudes as a
function of time for our determined pole orientation.  From this plot,
we can see that dramatic seasonal effects should be present around the
time of perihelion.  At the start of August~2001, the primary jet was
pointed nearly straight toward the Sun, so it should be expected that
the production rates peaked between this time and perihelion (taking
into account the trade-off between increasing solar radiation and
decreasing altitude of the Sun as seen from the jet).  Throughout
September, October and November, the primary jet receives less and
less sunlight, until, in early December, the Sun sets completely as
seen from the primary source.  The reduced level of sunlight during
this time manifests itself in the fading of the jet, and of the comet
in general, as shown in the luminosities in Figure~\ref{radprof4}.  We
note that by December, the secondary source may be starting to
activate, because the anti-jet profile ({\it i.e.}, the direction of
the secondary jet) is about twice as bright as the other directions.

Examination of our February and May images in light of our pole
position confirms the fact that the secondary jet must be on the
opposite hemisphere from the primary (in February, the projected north
pole lies at PA 158$^\circ$ and in May it lies at 130$^\circ$).
Although the secondary jet PA is within a few degrees of the rotation
axis PA in each case, our lack of temporal data means we cannot
evaluate any projection effects (or lack of them) that might indicate
how close the jet lies to the pole.  On the other hand, as discussed
in section~\ref{sec:dust}, the profile of the center of the jet in the
February image maintains an $m=-1$ slope out to distances beyond
30,000~km, which suggests that the active area must be close enough to
the pole that it is illuminated almost continuously during the
rotation of the nucleus.  From Figure~\ref{subearth}, we see that the
Sun was at a cometocentric latitude of $-30^\circ$ in February and
$-50^\circ$ in May, so the secondary active region is likely to be
situated within $\sim30-40^\circ$ of the pole, which would allow it to
receive nearly constant illumination during these times.

With the knowledge that the two jets are located on opposite
hemispheres and an understanding of when each source is illuminated by
the sun, we can now use the luminosity information from
Fig.~\ref{radprof4} to estimate the relative sizes of the two sources.
If we assume that the coma brightness in September is due to emission
only from the primary jet and that the brightness in February is due only
to emission from the secondary, then the ratio of brightness,
corrected by the solar illuminance, gives a zeroth order approximation
of the relative sizes of the active regions.  (Other factors, such as
the altitude of the sun as seen from the active area, the fraction of
time the sources are in sunlight, and differences in the material
emitted from each source will also affect the brightness, but for this
zeroth order computation, we neglect these effects.)  The luminosity
in September is a factor of 10 greater than in February, of which a
factor of about 2.5 can be attributed to the fall-off in solar
radiation.  Thus, the primary active area must be about four times
larger than the secondary.  Schleicher {\it et al.}~(2002) found that
the primary source has an area of 3.5~km$^2$ (4\% of the nucleus' surface
area), which means that the secondary source has an area around 1~km$^2$
or 1\% of the surface area. 

Returning to the issue of the pole position, we should consider the
results of the two previous attempts to determine the spin axis
orientation of comet Borrelly.  In 1997, Fulle~{\it et al.}~(1997) 
presented a solution derived from
modeling 20 images from the 1994 apparition.  Their results required
that the spin axis precess at an angle of 50$^\circ$, with a period of
about 2.5~years, to match the jet's appearance.  However, as discussed
earlier, we believe the nucleus must be near a state of pure spin,
with little or no precession.  To investigate the discrepancy between
our result and that of Fulle~{\it et al.}, we applied our pole
solution technique to their measurements ($PA_{SS}$ from their
Table~1).  The resulting set of pole/LOS planes are shown in the
bottom panel of Figure~\ref{sekpole}.  As can be seen, 14 of the 20
curves had intersections that were concentrated at the position
$\alpha=215^\circ,~\delta =-7^\circ$ with a scatter of about
4$^\circ$.  Of the six discrepant points, two have intersections well
away from those of the other curves, indicating that they are probably
misidentifications of the primary jet.  One is from early in the
apparition; the other is from very late, when the active region is not
illuminated. It is likely that this latter point was actually a
measurement of residual material similar to the protrusion we saw in
our February images.  The other four points, which are only marginally
discrepant, are from late in the apparition when the jet was not well
defined.  (These observations correspond to the December time frame of
the 2001 apparition, where we assigned uncertainties of 5$^\circ$ to
our PA measurements.)  Based on the images in the Fulle~{\it et al.}
paper, we believe that their measurements from later in the apparition
have errors larger than the global uncertainty of 1$^\circ$ that is
quoted for all of the measurements.  If we assume their uncertainties
are comparable to our December results and adopt errors on the order
of 3-4$^\circ$ for the later measurements, then even the marginally
discrepant measurements become consistent with the intersection of the
other 14 curves.  Thus, we conclude that our pole solution is robust
for both the 1994 and 2001 apparitions and that there is no need to
invoke precession or complex rotation to explain the 1994
measurements.  As one final comparison, we note that our predicted
pole PA of 94$^\circ$ for 11~November~1994 matches the direction of
the jet as seen in a contour plot of the HST image from that date
(Lamy {\it et al.}~1998).   

Sekanina (1979)  
presented another pole solution,
$\alpha=70^\circ,~\delta =-35^\circ$, that differs significantly from
our result.  To constrain his analysis, Sekanina used descriptions of
comet Borrelly from the four apparitions between 1911 and 1932.  He
assumed that the fan-shaped coma was produced by anisotropic emission,
and that the offset between the center of the fan and the sub-solar
point was produced by a thermal lag that, combined with the spin of
the nucleus, shifts the direction of the peak emission.  Using this
technique, Sekanina computed the pole position and a time-dependent
lag angle that best fit the observations from the four apparitions.
Based on our analysis of the rotation state, however, we know that
some of the basic assumptions behind Sekanina's model break down
because the active region is aligned with the axis.  Specifically, the
offset between the center of the fan and the sub-solar point is
produced by the relative directions of the sun and the jet throughout
the orbit, and is completely independent of the rotation rate of the
nucleus or the thermal lag in the sublimation rate.  Thus, Sekanina's
technique is not applicable in the case of comet Borrelly and the
solution that he found is not representative of the actual pole
position.

It can be argued that the rotation state of Borrelly's nucleus changed
drastically between 1932 and 1994, and that Sekanina's assumptions
were valid for the dates he used to constrain his models.  A change
this dramatic is unlikely to have occurred, however, because the
comet's nongravitational accelerations have remained nearly constant
since it was discovered (Yeomans 1971, Marsden
1999).   
The forces causing these accelerations
are produced by jets on the nucleus, so changes in the
rotation state or the location of the active areas will be reflected
in the nongravitational force terms.  Since the variations in the
nongravitational acceleration terms varied by only 12\% between
1904 and 2001, we conclude that the pole orientation and active area
locations have remained nearly the same throughout this time period.

Using our pole position, we computed the expected direction of the jet
on the dates the comet was observed between 1911 and 1932 (Sekanina
1979 and references therein).   
This allowed us to
evaluate whether our pole position could be used to predict the jet
direction over many apparitions.  In general, our predictions matched
the data in more cases than the model used by Sekanina, but a large
number of points still had major deviations from the observations.

To explore the issue further, we applied our pole solution technique
to the measurements from 1911, 1918 and 1925.  (Only one measurement
is given for 1932, so it is an indeterminate case.)  The results are
shown in Figure~\ref{sekpole}, with our 2001 pole solution represented
by the dot for comparison.  The 1911 data, which represent the
apparition with the best observing conditions, show a convergence of
curves at $\alpha=214^\circ,~\delta =+2^\circ$ (note that the one
discrepant curve was obtained under extremely poor observing
conditions, so we can discount it in the fit).  On the other hand, the
1918 and 1925 plots show no consistent set of intersections that would
indicate a preferred pole direction.

Unfortunately, we cannot thoroughly investigate the absence of a
single solution for these apparitions because of a lack of details
about the original position angle measurements.  The PA measurements
are less than ideal for our purposes, consisting only of short
descriptions of visual observations (Sekanina 1979 and references
therein).  These descriptions are often vague about what exactly is
being measured ({\it i.e.}, some entries simply state that there is an
elongation of the inner coma), which makes it difficult to evaluate
whether the listed PA refers to the primary jet or not.  Also, factors
such as radiation pressure can introduce asymmetries or curvatures in
the jet, which can affect the apparent position of the jet center.
With CCD images, we have the ability to process and enhance the images
to detect these effects and correct for them, if necessary, something
not possible with visual observations.  Another problem is that,
although the measurements usually appear to be given to the nearest
5$^\circ$ increment, there is no mention of their uncertainties, so it
is not obvious which PAs can be considered accurate and which may be
suspect.  Furthermore, there is no discussion of the techniques used
to measure the position angles, which raises the possibility that
systematic offsets could also be present.  

The observing geometries that were present in 1911, 1918, 1925 and
1932 suggest that features in the coma would be more difficult
to resolve on each successive apparition, thus contributing to larger
measurement uncertainties.  The observing geometry was best in 1911,
with the comet closest to the Earth.  This gave the highest resolution
and presumably the best contrast of the jet against the background.
For the 1911 apparition, the pole/LOS curves converge on a position very
close to our 2001 solution.  (Again, we know the discrepant curve was
obtained under poor observing conditions, so we assume it has large
measurement errors and discount it.)  For the 1918 and 1925
apparitions, the lack of consistent solutions might be attributed to
uncertainties that are generally larger than those seen in 1911,
resulting from the decreasing spatial resolution and lower contrast on
each apparition.  Indeed, by 1932, the conditions had deteriorated to
the point that the jet was only detected on one date, even though
photographic plates were being used for observations at that time.
Given the potential problems with the observations, we can accept the
1911 pole position as being fairly consistent with our 2001 pole
solution.  Similarly, the 1918 and 1925 curves pass close to our
solution, so with large enough error bars (generally 5-10$^\circ$),
most of these curves would be consistent with our solution, as well.

Alternatively, if we accept the PA measurements as quoted, we see an
interesting trend.  Looking at each curve from the 1911-1925
apparitions, we note that, in almost every case, the point of closest
approach to our solution lies at a declination north of our pole.  In
fact, out of all the measurements, only two curves lie to the south of
our solution, one from 1911 and the other from 1918 (and the 1911
curve is from the PA measured under poor observing conditions).  If
the PAs had random errors, then we should expect that an equal number
of curves would lie to the north and south, which is clearly not the
case.  This trend suggests that either there is a systematic error
that preferentially shifts the curves north, or else the pole was
pointed at a declination 5-10$^\circ$ north of our 2001 position when
these observations were made.  If the latter trend is the case, then
we must conclude that the pole has shifted its orientation over the
past 70 years.  Additional evidence for a shift in the pole position
is presented in Section~\ref{sec:density}. 

For the most part, the nongravitational parameters listed by
Marsden~(1999) show that $A_2$ has shifted in at least three small
stages, rather than in a single jump.  This suggests that the
pole has been migrating gradually with time, possibly due to small
torques produced by the secondary jet, rather than jumping in a single
apparition as might be produced by an impact.  If we adopt the pole
position found from the 1911 data ($\alpha=214^\circ,~\delta
=+2^\circ$) and assume a constant drift in position, we find that the
pole would have moved $\sim$7$^\circ$ in the past 13 orbits, or about
0.5$^\circ$ per orbit.  This level of change is too small to be
detectable from one apparition to the next, and so would not conflict
with any of the assumptions that we adopted in our analyses.  If this
gradual migration of Borrelly's pole proves to be real, it could be
the first clear example of the long-term evolution of cometary spin
axes discussed by Samarasinha~(2002), in which cumulative effects of
collimated outgassing cause the direction of a comet's spin axis to
spiral towards the direction of the comet's perihelion (or
aphelion). 
We also note that Schleicher {\it
et al.}~(2002) see similar evidence for the pole shift in their models
of the jet morphology and they discuss its implications in more
detail.

\subsection{Mass and Density of the Nucleus}\label{sec:density}

Borrelly is unusual in that it exhibits nongravitational force
coefficients that differ significantly from zero, yet they have
remained nearly constant since the comet's discovery (Yeomans 1971).   
In most comets, an active region large enough to
accelerate the entire nucleus will also produce torques that alter its
rotational properties, which in turn affects the future
nongravitational forces.  In the case of Borrelly, however, the force
from the primary jet is directed along the spin axis, so no torque is
generated to introduce precession or to change the spin rate of the
comet.  Thus, the rotational state remains unchanged and the
nongravitational forces are essentially repeatable from one apparition
to the next.  (The possible drift in the pole direction will be
addressed later.)  Taking advantage of the known nongravitational
accelerations, we utilized our pole orientation in an analysis to
estimate the mass and density of the nucleus.

The primary observational manifestation of the nongravitational forces
is an advance or delay in the time of perihelion passage, effectively
changing the comet's orbital period.  (This effect is reflected in the
nongravitational coefficient, $A_2$.)  Rickman~(1989) showed that the
period change could be related to the acceleration, ${\bf j}$, in the
orbital plane:
\begin{equation}\label{eq:deltap}
\Delta T = \frac{6\pi (1-e^2)^{0.5}}{n^2} \left( \frac{e}{p} \int^T_0
  j_r \sin{\theta}\, dt +   \int^T_0 \frac{j_t}{R_h}dt \right)
\end{equation}
where $t$ is time, $T$ is the orbital period, $e$, $n$ and $p$ are the
orbital eccentricity, mean motion and semi-latus rectum and $\theta$
is the true anomaly.  $j_r$ and $j_t$ are the radial and transverse
nongravitational accelerations.  The integrals reflect the fact that
the period change is the result of the net sum of the accelerations
throughout the entire orbit.  The force in the orbital plane, ${\bf
F}$, which is related to the mass of the nucleus, $M$, by ${\bf
j}={\bf F}/M$, is produced by the directed outflow of the gas and dust
\begin{equation} \label{eq:force}
{\bf F} = - \sum_i Q_i m_i{\bf v}_i.
\end{equation}
The $Q_i$ are the production rates of the different species of masses
$m_i$, which have emission velocities~${\bf v}_i$.

Given this result, it is clear that with measurements of the
production rates and period change and a good understanding of the gas
outflow characteristics, it is possible to determine the mass of the
nucleus.  Furthermore, if the dimensions of the nucleus are known, then its
bulk density can be found.  Although this technique has been used in a
statistical sense to study the effects of nongravitational forces
({\it e.g.,} Rickman {\it et al.}~1987, Sekanina~1993), there are usually
too many unknowns to apply it to specific comets.  For example, the
directionality of the forces cannot be determined unless the rotation
state and locations of the active areas are known.  Even if these
values have been determined from other information, however, the combined
effects of rotation and thermal lag (which are not well understood)
introduce further complications that must be addressed.  Much work has
been done with comet Halley ({\it e.g.}, Rickman~1989, Sagdeev~1988),
but the uncertain rotation state and the effects of thermal lag
prevent strong constraints on the mass and density.  In contrast, most
of the emission from comet Borrelly was directed along the rotation
pole, whose direction is fixed and known.  This alignment also means
that the nucleus rotation and thermal lag have no effect on the
direction vector of the nongravitational forces.  Thus, we have the
opportunity to set limits on the mass and density of Borrelly's
nucleus that could be the most well-constrained of any comet to date.

To simplify our analysis, we made two important assumptions: First, we
assumed water was the only non-negligible mass loss component.  This is
justified because the water production is much greater than that for
any other gas species ({\it e.g.}, Schleicher {\it et al.}~2002), with
all other species combined contributing at a level of only 10--20\%
that of water (Rickman~1989 and references therein).  Second, we
assumed that all of the mass loss came from the primary jet and was
directed along the direction of the rotation pole.  This is also
justified because we know from the luminosities in
Figure~\ref{radprof4} that emission from the primary jet is about an
order of magnitude greater than from the secondary jet.  Furthermore,
not only do Schleicher {\it et al.} conclude that 90--100\% of the
water production comes from the primary jet, but the applied force from
this jet peaks near perihelion, where a given acceleration will
produce a larger $\Delta T$ than accelerations applied at larger
heliocentric distances.  This combination of factors means that, to
first order, the nongravitational contribution from the secondary jet
can be neglected.  As for any isotropic emission, we know that it is
probably also small relative to that from the primary jet, and
therefore the acceleration that it produces can be neglected compared
to the highly directed emission from the jet.  Even if the fraction of
gas production due to the isotropic component is larger than we
expect, the acceleration produced by isotropic outflow will, to first
order, mimic that from the jet ({\it i.e.}, the pole is directed
toward the sun near perihelion, so only the sunward hemisphere will be
active and the net force will be in the same general direction as that
from the jet).

Returning to equation~\ref{eq:force}, we can evaluate each of the
components that comprise the nongravitational force.  Because the
direction of the jet in inertial space is known, it is a simple matter
to compute the directional components of the emission velocity, ${\bf
v}$, as a function of time.  To simplify this computation, we
converted the direction of the pole from right ascension and
declination to coordinates in the orbit reference frame ($I_p,L_p$).
We defined the obliquity of the pole, $I_p$, to be the angle of the
pole relative to the orbital angular momentum vector.  The orbital
longitude of the pole, $L_p$, is measured from the anti-solar
direction at perihelion and increases in the direction of the comet's
motion.  (In these coordinates, the pole is oriented at
$I_p=102^\circ, L_p=145^\circ$.)  The magnitude of the emission
velocity projected into the orbital plane is then simply $v\sin I_p$,
where $v=|{\bf v}|$.  This can be separated further into the radial
and transverse components: $v_r = v\sin I_p \cos(\theta - L_p)$ and
$v_t = v\sin I_p \sin(\theta - L_p)$, respectively.  

For the emission velocity of water, we adopted the relation used by 
Rickman~(1989) in his analysis of comet Halley:
\begin{equation}
   v = \eta^* \left(\frac{1}{1-\alpha}\right)v_{therm}
    \label{eq:velocity}
\end{equation}
where $\eta^*$ is a dimensionless factor dependent on the Mach number
of the flow above the nonequilibrium boundary layer; $\alpha$ is the
fractional recoil flux of gas molecules that have been turned around
and return to impact on the surface of the nucleus, producing an
increase in the momentum transfer; $v_{therm} = (8kT/\pi m)^{1/2}$ is
the thermal gas velocity; $k$ is Boltzmann's constant; $T$ is the
temperature; and $m$ is the mean molecular mass.  For a temperature of
200~K, the thermal velocity $v_{therm}$ is about 500~m\,s$^{-1}$, which we
adopted for our analysis.  We then used values of $\eta^*=0.5$ and
$\alpha=0.25$ (Wallis and Macpherson 1981, Rickman~1989, Peale~1989)
to obtain an emission velocity for the gas molecules of
$v$=330~m\,s$^{-1}$.  We recognize that there are a number of
uncertainties in this representation, both in the physics of the gas
outflow and in the true values of the variables, and this will be
addressed later.

To represent the mass loss rate, we need the water production rates
from comet Borrelly as a function of time.  Schleicher {\it et
al.}~(2002) modeled this water production using a vaporization model
that includes dependences on both heliocentric distance and incidence
angle of the sunlight.  The model was constrained using measurements
of the water production, based on their narrow band photometry of OH.
Unfortunately, the water production rates prior to --50~day are not
constrained by observations because the comet was in solar conjunction
during this time frame.  Because the primary jet was in sunlight
starting around --200~days, we were concerned that errors in the
production rates from --200 to --50~days would affect our results.  To
investigate this issue, we performed a series of tests to determine
how the density changes if the production rates for the time period
--200~to --50~days are altered.  As it turns out, the final result is
fairly insensitive to this time frame, because the peak production is
close to perihelion, where the nongravitational forces are most
efficient.  In the most extreme of our tests, we turned the water
production completely off until --50 days, at which time it was
``turned on'' at the level computed by Schleicher {\it et al.}  Even
with this dramatic change, the final density shifted by less than
10\%, which is well within other uncertainties discussed below.  Given
this result, we adopted the Schleicher {\it et al.} production rates,
as given, for our analysis.

Recalling the relation ${\bf j}={\bf F}/M$ and replacing the above
expressions for the different terms in equations~\ref{eq:deltap}
and~\ref{eq:force}, we can solve for the comet's mass:
\begin{eqnarray} \label{eq:mass}
  M = 1.26\times10^{-10}\,\frac{(1-e^2)^{0.5}}{n^2 \Delta T} \,
       m_{H_2O}\,  v\, ~
      \left[ \frac{e}{p} \int^T_0  Q_{H_2O}\, \sin I_p \,
      \cos\,(\theta - L_p)\, \sin{\theta}\, dt
       \right. \nonumber  \\
      \left. +~\int^T_0  Q_{H_2O}\, \frac{\sin I_p \, 
      \sin\,(\theta - L_p)\,}{R_h}\,dt 
       \right]
\end{eqnarray}
where $n$ and $Q_{H_2O}$ are expressed in sec$^{-1}$, $T$ and
$\Delta T$ in days, $m_{H_2O}$ in g, $v$ in m\,sec$^{-1}$, $p$ and
$R_h$ in AU and $M$ in g.  Thus, by determining the change in the
orbital period from the nongravitational forces and integrating the
acceleration over the entire orbit (or at least over the times that
the primary jet is active) we can determine the mass of the nucleus.

Using the nongravitational force coefficient, $A_2 = -0.0376$
(M.P.C.~31664), and the procedures outlined by Rickman {\it et al.}
(1987), we computed the change in the period of $\Delta T =
-0.052$~day for comet Borrelly.  With this value, we found a mass of
the nucleus of $1.8\times10^{16}$~g.  From the DS1 results
(Soderblom {\it et al.}  2001), the dimensions of the nucleus are
$4\times4\times8$~km.  If we assume the nucleus can be represented by
a triaxial ellipsoid, then it has a volume of 67~km$^3$, which leads
to a bulk density of 0.27~g\,cm$^{-3}$ for the nucleus.

Given the fortuitous alignment of the primary jet and the rotation
axis, the largest source of error in our computations comes from the
uncertainties in the momentum transfer between the ejected material
and the nucleus.  This encompasses a number of physical mechanisms,
including the sublimation of the ices, the hydrodynamics of the gas
flow, and the role of dust in the scenario (Peale 1989, Skorov and
Rickman 1999).  The effect of all these mechanisms tends to be
concentrated into a single parameter in our analysis -- the average
gas velocity, $v$.  So, by estimating the range of acceptable
velocities that could result from comprehensive gas flow calculations,
we can constrain the total range of densities that would result.
Conveniently, the mass and density both vary linearly with the
velocity, making the variations trivial to compute.

There are no measurements of the gas flow very near the nucleus, which
is the region of interest in the nongravitational acceleration
analysis.  However, during the Halley spacecraft encounters, an {\it
in situ} measurement showed a gas velocity that would correspond to
850\,m sec$^{-1}$ at 1~AU (Krankowsky {\it et al.}~1986).  We consider
this to be an extreme upper limit to the velocity for several reasons:
The measurement was obtained well beyond the boundary layer, where
acceleration should have increased the average velocity (Wallis and
Macpherson 1981); Halley was much more active than Borrelly, which may
have contributed to higher velocities (Combi 1989); and Halley was
much closer to the sun (0.89~AU) at the time of the measurement than
Borrelly ever gets ($q$=1.36~AU), so any inverse $r$-dependence would
mean that Borrelly would have a lower velocity.  At the other extreme,
Crifo (1991) pointed out that gas flowing from the nucleus reaches the
transition to a sonic flow within the first few meters from the
nucleus' surface.  Combi (1989) stated that the initial outflow speed
at the sonic point is of order 300\,m sec$^{-1}$, so the gas outflow
velocity at 1\,{\sc au} is unlikely to be below 300\,m sec$^{-1}$.  We
adopt this as our lower limit.  By using the range $300 < v_{therm} <
850$~m~sec$^{-1}$, and replacing the respective values in place of the
thermal velocity in Eq.~\ref{eq:velocity}, we find that the density
has a range $0.16 < \rho < 0.46$~g~cm$^{-3}$.  Given that the limits
on the velocity are believed to be the most extreme acceptable, these
should be considered 3$\sigma$ limits.

Skorov and Rickman~(1999) used more detailed hydrodynamic models to
show that the models implemented by Rickman~(1989), which were adopted here,
underestimate the momentum transfer that produces the nongravitational
acceleration.  In order to account for this problem, they suggested an
average multiplicative factor of 1.8 as a correction to the density.
Applying this to our results, we obtained a mass of
3.3$\times10^{16}$~g and density of 0.49~g~cm$^{-3}$, and the range of
possible densities is $0.29 < \rho < 0.83$~g~cm$^{-3}$.

As was mentioned earlier, the dust probably contributes to the
non-gravitational acceleration, but has not been taken into account in
any of the available models (though Skorov and Rickman (1999)
acknowledged that it should be considered).  Strictly speaking, all of
the momentum in the system comes from sublimation of the gas and the
dust motions merely reflect momentum that has been transferred from
the gas.  In principle, then, using the gas production rates and the
gas dynamics at the surface of the nucleus should be sufficient to solve
the momentum equations.  In practice, however, there are two
mechanisms involving the dust that can alter the momentum balance.
First, if a gas molecule is emitted from the nucleus, strikes a dust
grain and reflects back to the nucleus (in the same manner as the
recoil force in the gas flow), then the gas molecule is acting to
transfer momentum from the dust to the nucleus.  This mechanism would
only be efficient close to the surface, where large numbers of
reflected molecules would intersect the nucleus.  Second, if a gas
molecule is emitted from the nucleus and sticks to a dust grain, then
it transfers its momentum to the dust, while at the same time
effectively removing itself from the observable coma.  This means that
there is momentum in the dust that is not accounted for in the
measurement of the gas production rates.

Due to the action of these two mechanisms, the dust is involved in the
total momentum transfer and a comprehensive analysis would need to
take this into account.  Unfortunately, the physics of the dust/gas
flow are not well understood at present and there are too many
variables to provide any significant constraints on the dust
contribution to the nongravitational forces.  Among the questions that
need answering are: Where does the dust acceleration take place? What
is the scattering efficiency in a dust/gas collision? What is the
dust to gas ratio of the comet? What is the composition and structure
of the dust? How does the presence of the dust affect the gas
flow?  Presumably, the effects of the dust might cause the computed
density to rise by a factor of 5-10\% or higher, though the exact
contribution will remain unknown until better hydrodynamic models are
developed to address the issue.

The low bulk density that we found in our analysis indicates that the
nucleus must be fairly porous, even if it is composed primarily of
ices.  Formation models of porous bodies ({\it e.g.}, Donn and Duva
1994, Donn 1990)
show that, for low
density material, even low-velocity impacts will compress and heat the
material in the impact zone, producing changes in the structure.
Given the low average density of Borrelly, we can conclude that the
accretion processes that formed the nucleus must have occurred with
fairly low relative velocities ($<5$~m\,sec$^{-1}$).  Furthermore, the
nucleus probably doesn't have a homogeneous structure, because even
low velocity impacts encountered during the comet's formation would
alter the density in the collision zones, while other regions remain
unaffected.

As discussed earlier, the nongravitational accelerations have remained
nearly constant since the start of the century, varying by only about
12\% between 1911 and 2001 (Marsden 1999).  However, this small change
in $A_2$ is significant enough that we believe it provides another
line of evidence supporting the conjecture that the pole has changed
position over time.  The changing nongravitational acceleration means
that one or more of the factors producing the acceleration (the water
production rate, the direction of the force, and the mass of the
nucleus) has changed.  We can rule out the possibility that a
reduction in the mass of the nucleus is the cause, because Schleicher
{\it et al.}~(2002) compute the mass loss from water sublimation to be
about 10$^{13}$~g per orbit.  If the production rates have remained
similar over the past 13 orbits, then this amounts to a total mass
loss of less than 1\% since 1911.  No measurements exist of the water
production rates in the early part of the century, so we cannot rule
out changes in the water production, but the results from
Section~\ref{section:pole} provide us with the opportunity to
investigate whether the pole might have changed its orientation over
time.  For this analysis, we expect that the reaction force, the
direction in which it is acting, and the observed nongravitational
acceleration should combine in such a way that we always compute the
same density for the comet.  Thus, we use the density that we computed
for 2001, $\rho=0.49$~g~cm$^{-3}$, as our comparison value and explore
how changes in the pole position affect the result.

First, we examined the case in which the direction of the
nongravitational force was the same for 1911 as it was in 2002 (e.g.,
what density would be computed if the pole didn't change position).
The nongravitational force coefficient for 1911 was given by Marsden
(1999) as $A_2=-0.0421$.  With this value and the 1911 orbital
elements from the same source, we found that $\Delta T = -0.059$~day.
We then used our 2001 pole solution and 2001 production rates to
integrate the nongravitational forces.  (The 1911 and 2001 orbits are
similar enough that the production rates would be essentially the same
for the same pole orientation.)  With this configuration, we computed
a density for Borrelly of 0.43~g~cm$^{-3}$ (including the 1.8 scaling
factor), which differs from our 2001 solution by 12\%.  This result
simply reflects the fact that if all other factors are constant, then
an increase in $A_2$ will produce a corresponding decrease in the
density. 

Next, we looked at the nongravitational acceleration that would result
if the pole had changed position between 1911 and 2001.  For this
case, we used the 1911 pole solution discussed in
section~\ref{section:pole} ($\alpha=214^\circ,~\delta =+2^\circ$) as
the direction of the reaction force.  Because the pole position is
different, the water production rates will differ as well.  To keep
our test internally consistent, we used water production rates that
were computed for the 1911 apparition by D. Schleicher (private
communication) in the same manner that he used to model the 2001 water
production.  (We note that the production rates were computed for the
1911 pole position found by Schleicher {\it et al.} (2002), but their
solution differs by only a couple degrees from ours, and so the
production rates should not differ enough to significantly affect our
results.)  Using these parameters, along with the 1911 value for
$A_2$, we computed a density of 0.49~g~cm$^{-3}$, which is essentially
identical to our 2001 result.  This agreement shows that the pole
solution we found from the 1911 data is consistent with the
nongravitational forces that were measured for that time frame.
Although this is not conclusive proof for a shift in the pole
position, it does provide a clean explanation for the difference in
the nongravitational forces between 1911 and 2001, and thus supports
the conjecture that the pole shift might be real.

\section{Summary}

We obtained imaging and spectroscopic data on comet 19P/Borrelly at
the time of the Deep Space~1 flyby in September 2001 and in subsequent
months.  These observations help to place the DS1 encounter data into
a more global view.  The DS1 images confirm our picture of a comet
with a strong, narrow jet along the waist and yield the dimensions of
the nucleus.

From our observations, we have drawn the following conclusions:

\begin{itemize}
\item We utilized the nongravitational accelerations of comet Borrelly
to compute a mass of the nucleus of $3.3\times10^{16}$~g and a density
of 0.49~g\,cm$^{-3}$ (with a range of $0.29 < \rho <
0.83$~g\,cm$^{-3}$).  Because the direction of the reaction force and
the water production rates are both well-known (and highly repeatable
from one apparition to the next), and because the dimensions of the
nucleus were measured in situ, this is the least model-dependent comet
density known to date.

\item The strong jet seen in the DS1 images that emanates from the
waist of the comet is aligned with the comet's rotation axis.  We
determined the orientation of the pole to be $\alpha=214^\circ,~\delta =
-5^\circ$, with an uncertainty of 4$^\circ$, which is consistent with
other solutions, including the DS1 estimate.  Given this orientation,
the jet was pointed about 40$^\circ$ from the Sun at the time of the
DS1 encounter.  There is also evidence that the pole orientation
changed by 5-10$^\circ$ between the 1911 and 1994 apparitions.

\item The position of the pole results in a strong seasonal effect
in the activity levels of the jets.  As the comet receded from the
Sun, the primary jet at the pole received less and less illumination.
Eventually the primary jet turned off and a secondary, much weaker
jet, turned on.  The secondary jet is located on the opposite
hemisphere from the primary jet and probably lies within 30-40$^\circ$ of
the pole.

\item The distribution of the gas and dust in the coma is quite
asymmetric in the sunward/anti-sunward directions.  However,
perpendicular to this direction, the gas seems to be quite
symmetrically distributed.  The distribution of C$_{2}$ gas in the
sunward direction in November 2001 is quite uniform with cometocentric
distance out to 50,000\,km.  Such a distribution cannot be easily
reproduced with simple two-component models.

\item A comparison of the C$_{2}$ and CN gas distributions in the coma
on September 2001 and November 1994 shows a remarkable similarity.
Except for the geocentric distance, the viewing geometries from these
dates were nearly identical.  The comet shows the same asymmetries in
both apparitions and the gas column densities are the same.  This
points to a very stable gas production and is another piece of
evidence that the comet must be in simple rotation.

\item The comet is mildly depleted in C$_{2}$ and C$_{3}$ relative to CN.

\item The dust in the coma is very red, with the tailward region being
much redder than the sunward jet. This suggests that the particles in
the primary jet are, on average, smaller than those in the rest of the
coma and tail.  However, residual particles from the primary jet are
still seen in February, which indicates that the particle size
distribution, even in the primary jet, contains many large grains.
The jet appears to exhibit a steady-state outflow while the tail and
perpendicular regions show evidence for radiation pressure acting on
the dust.

\end{itemize}
\vspace{0.5in}
\begin{center}Acknowledgements\end{center}
This research was supported by NASA Grants NAG5-9003 and NAG5-4384.
We thank Drs. David Schleicher, Laura Woodney and Nalin Samarasinha
for helpful discussions, Dr. Laurence Soderblom for communicating
the Deep Space 1 pole solution to us prior to publication, and
Dr. Beatrice Mueller for her comments on the manuscript.

\clearpage
\newpage
\begin{table}

\caption[obslog]{Observing Parameters}\label{log}
\centering

\begin{tabular}{r@{ }l@{ }lr@{ -- }l@{\ \ \ }ccr@{.}lccp{1.5in}}
\multicolumn{12}{l}{\bf Spectroscopy} \\
\hline
 \\ [-10pt]
 & & & \multicolumn{2}{c}{ } & R$_h$ & $\Delta$ & 
\multicolumn{2}{c}{$\dot{\rm R}_h$}  & PA & PA &  \\
\multicolumn{3}{c}{Date} & \multicolumn{2}{c}{UT Range} &
({\sc au}) & ({\sc au}) &
\multicolumn{2}{c}{(km\,sec$^{-1}$)} &
 Sun$^\dagger$ & Slit$^\dagger$ &
Comments \\
\hline
22 & Sep & 2001 & 10:41 & 11:50 & 1.36 & 1.48 & \mbox{}\hspace{1em}+1&3 & 100.7 & 90 & non-photometric\\
23 & Sep & 2001 & 10:38 & 11:44 & 1.36 & 1.47 &+1&4 & 101.2 & 0 & photometric\\
25 & Sep & 2001 & 10:41 & 11:46 & 1.36 & 1.46 &+1&8 & 102.1 & 0 & photometric\\ [5pt]
19 & Nov & 2001 & 11:00 & 12:05 & 1.55 & 1.32 &+9&2 & 117.5 & 90 & non-photometric\\
21 & Nov & 2001 & 10:12 & 11:15 & 1.56 & 1.31 &+9&4 & 117.6 & 117 & photometric\\
21 & Nov & 2001 & 10:24 & 12:08 & 1.56 & 1.31 &+9&4 & 117.6 & 207 & photometric\\
\hline
\end{tabular}

\vspace*{5pt}
\begin{tabular}{r@{ }l@{ }lr@{ -- }l@{\ \ \ }cccccp{1.4in}}
\multicolumn{11}{l}{\bf Imaging} \\
\hline
 \\ [-10pt]
 & & & \multicolumn{2}{c}{ } & R$_h$ &  $\Delta$  &\hspace*{4.7em} & PA & \\
\multicolumn{3}{c}{Date} & \multicolumn{2}{c}{UT Range} &
({\sc au}) &  ({\sc au}) & Phase & Sun$^\dagger$  & Filter & 
Comments \\
\hline
21 & Sep & 2001 & 11:14 & 11:36 & 1.36 & 1.48 & 41.1 & 100.2 & V,R & non-photometric \\
22 & Sep & 2001 & 11:10 & 12:02 & 1.36 & 1.48 & 41.2 & 100.7 & V,R & non-photometric \\
23 & Sep & 2001 & 10:54 & 11:22 & 1.36 & 1.47 & 41.3 & 101.2 & V,R & photometric \\[5pt]
12 & Nov & 2001 & 11:18 & 11:46 & 1.52 & 1.33 & 40.1 & 117.1 & V,R & photometric \\[5pt]
04 & Dec & 2001 & 09:07 & 12:50 & 1.64 & 1.30 & 37.0 & 116.6 & V,R & photometric \\
05 & Dec & 2001 & 11:29 & 12:49 & 1.64 & 1.29 & 36.8 & 116.4 & V,R & photometric \\
06 & Dec & 2001 & 10:55 & 12:39 & 1.65 & 1.29 & 36.7 & 116.2 & V,R & non-photometric \\
07 & Dec & 2001 & 11:32 & 12 05 & 1.66 & 1.29 & 36.5 & 116.1 & R & non-photometric \\[5pt]
07 & Feb & 2002 & 08:57 & 12:48 & 2.08 & 1.35 & 22.8 &  70.7 & V,R & photometric \\
08 & Feb & 2002 & 09:47 & 12:48 & 2.09 & 1.35 & 22.6 &  69.2 & V,R & photometric \\  [5pt]
17 & May & 2002 & 03:25 & 04:00 & 2.78 & 2.45 & 21.1 & 303.9 & R & non-photometric \\
18 & May & 2002 & 04:37 & 05:10 & 2.79 & 2.47 & 21.1 & 303.4 & V,R & non-photometric \\
\hline
\multicolumn{10}{l}{\hspace*{0.1in}$\dagger$ Position angle
measured North through East} \\
\end{tabular}

\end{table}

\begin{table}
\caption{Derived Haser Model Production Rates}\label{prodrates}
\centering
\begin{tabular}{lccc}
 \\ [-5pt]
\hline
 & CN & C$_{2}$ & C$_{3}$ \\
\multicolumn{1}{c}{Date} & log Q/v & log Q/v & log Q/v \\
 & (mol sec$^{-1}$) & (mol sec$^{-1}$) & (mol sec$^{-1}$) \\
\hline
23 Sep 2001 & 25.50$\pm0.01$ & 25.23$\pm0.01$ & 24.60$\pm0.02$ \\
21 Nov 2001$^\dagger$  & 24.98$\pm0.01$ & 24.85$\pm0.01$ & 24.16$\pm0.03$ \\
\hline
\multicolumn{4}{l}{Error bars are formal errors of the fits} \\
\multicolumn{4}{l}{$^\dagger$ Perpendicular to the Sun/anti-Sun line} \\
\end{tabular}
\end{table}

\clearpage
\newpage
\newpage
\begin{center}{\bf Figure Captions}\end{center}

\begin{figure}[p]
\vspace{8in}
\includegraphics{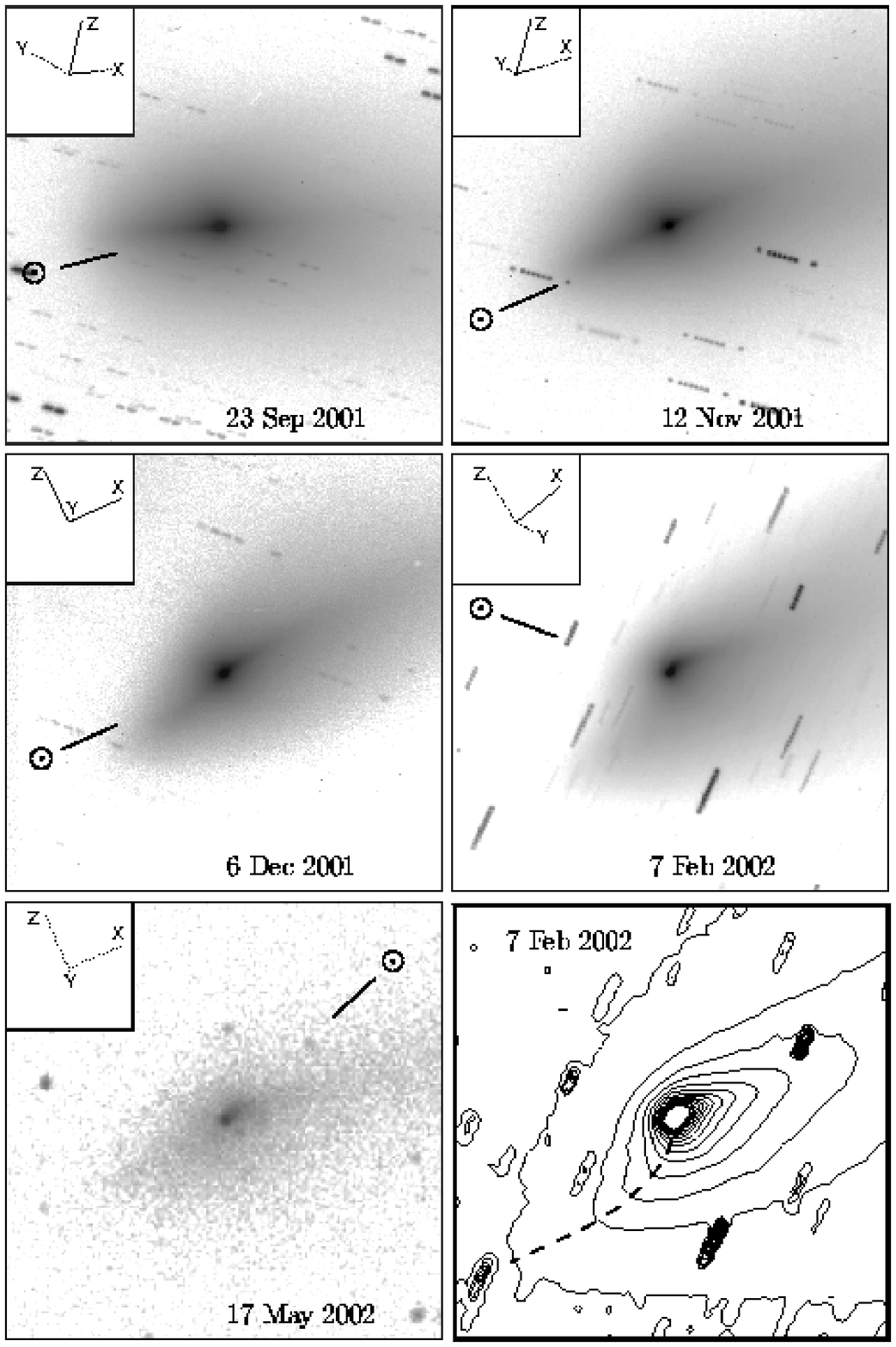}
\caption[fig1]{Farnham and Cochran}\label{images}
\end{figure}

\noindent
{\bf Figure~\ref{images}:} A sequence of 5 R-band images of comet
Borrelly, showing the evolution of the coma, and a contour plot of the
February data.  In each frame, North is at the top, East is to the
left and the field of view is $2.5\times10^{5}$~km.  The inset for
each image depicts an inertial coordinate system relative to the
comet's orbit, where the X~axis extends in the anti-solar direction at
perihelion, the Y axis is the velocity vector at perihelion and the Z
axis is parallel to the orbital angular momentum vector.  The length
of the axis in the inset indicates the amount of foreshortening, with
solid lines extending toward the Earth and dotted lines extending
away.

\begin{figure}[p]
\vspace{8in}
\includegraphics{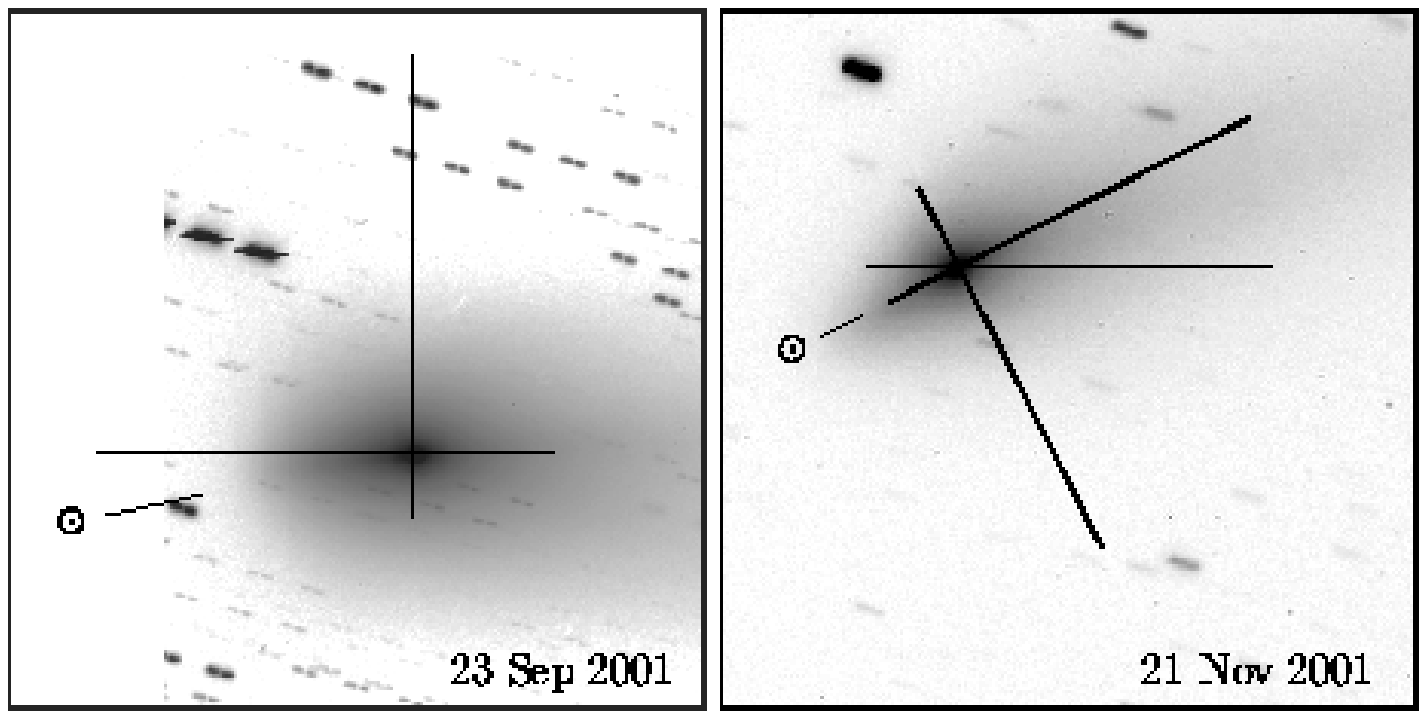}
\caption[fig2]{Farnham and Cochran}\label{imageslit}
\end{figure}

\noindent
{\bf Figure~\ref{imageslit}:} Images of comet Borrelly from
23~September and 20~November, showing the position of the spectrograph
slit for the different observing runs.  The orientation of the slit
for each particular night is listed in Table~I.  North is at the top,
East is to the left and the field of view of each image is
$3.3\times10^{5}$~km. The November image is courtesy of L.
Woodney (Personal communication).

\begin{figure}[p]
\vspace{8in}
\includegraphics{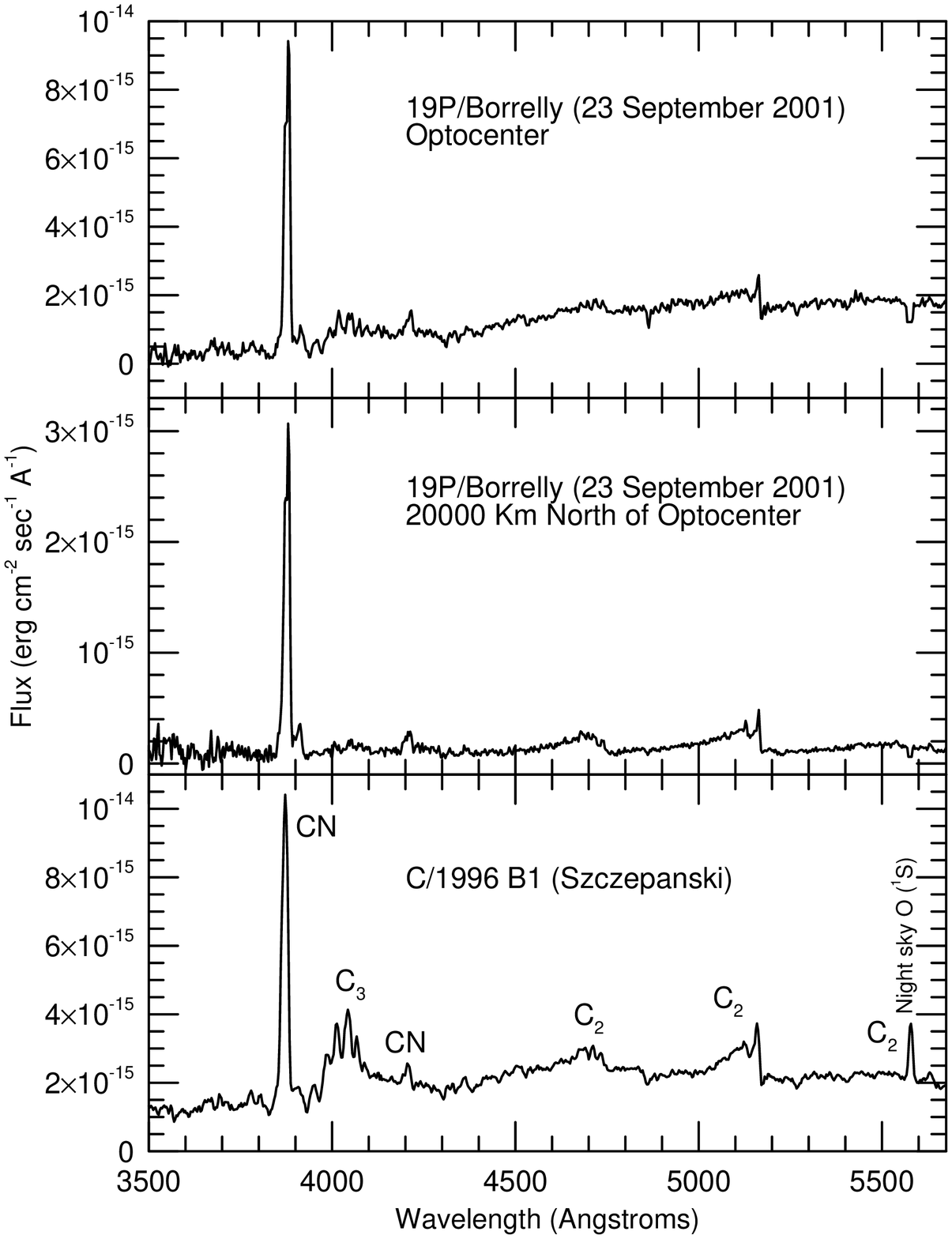}
\caption[fig3]{Farnham and Cochran}\label{spectrum}
\end{figure}

\noindent
{\bf Figure~\ref{spectrum}:} The spectrum of comet 19P/Borrelly on the
optocenter and 20,000\,km from the optocenter are compared with comet
C/1996 B1.  The Borrelly optocenter spectrum shows an enhanced
continuum over the off-optocenter spectrum.  For comparison, we show
an optocenter spectrum of comet C/1996 B1 (Szczepanski) (not sky
subtacted) when R$_h$=1.47, $\Delta=0.55$\,{\sc au}.  Borrelly does
not appear to have as much C$_{2}$ or C$_{3}$ relative to CN as does
Szczepanski.

\begin{figure}[p]
\vspace{8in}
\includegraphics{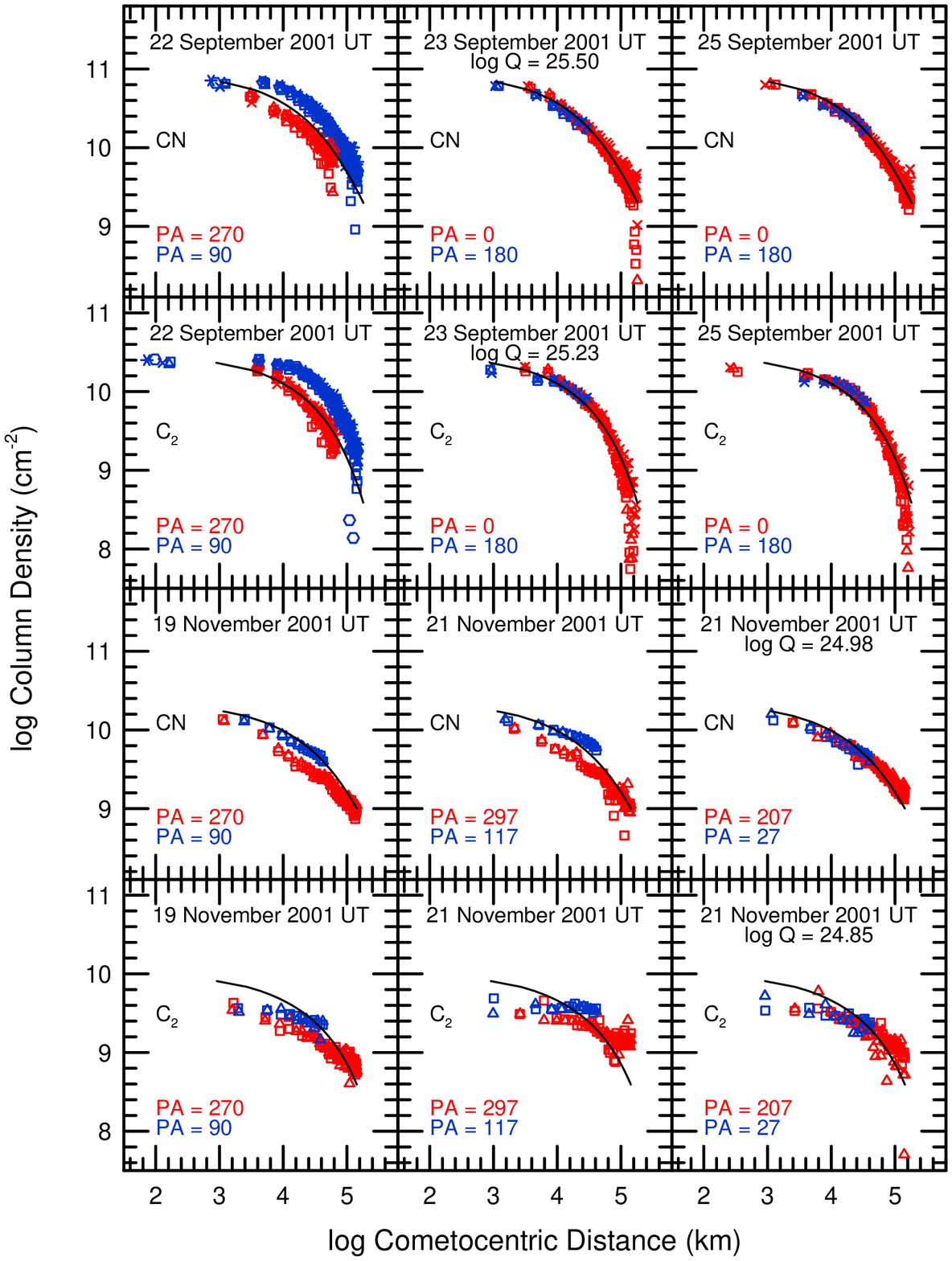}
\caption[fig4]{Farnham and Cochran}\label{gas}
\end{figure}

\noindent
{\bf Figure~\ref{gas}:} The distribution of the CN and C$_{2}$ gas for
the nights on which spectroscopic data were obtained.  The slit was
set to various position angles relative to the position angle of the
Sun (see Table~\ref{log} for Sun position angles).  On each night, at
least two spectral images centered on the optocenter were obtained and
each spectral image is plotted as a different symbol (on 22 September
there are 5 different spectral images).  Note the extremely good
agreement for different spectral images on the same night.  Color is
used to denote on which side of the optocenter each spectrum was
obtained.  The curve in each panel is a Haser model and is described
more fully in the text.  Note the high degree of asymmetry when the
slit was oriented along the Sun/anti-Sun line but the symmetry when
the slit was perpendicular to this line.

\begin{figure}[p]
\vspace{8in}
\includegraphics{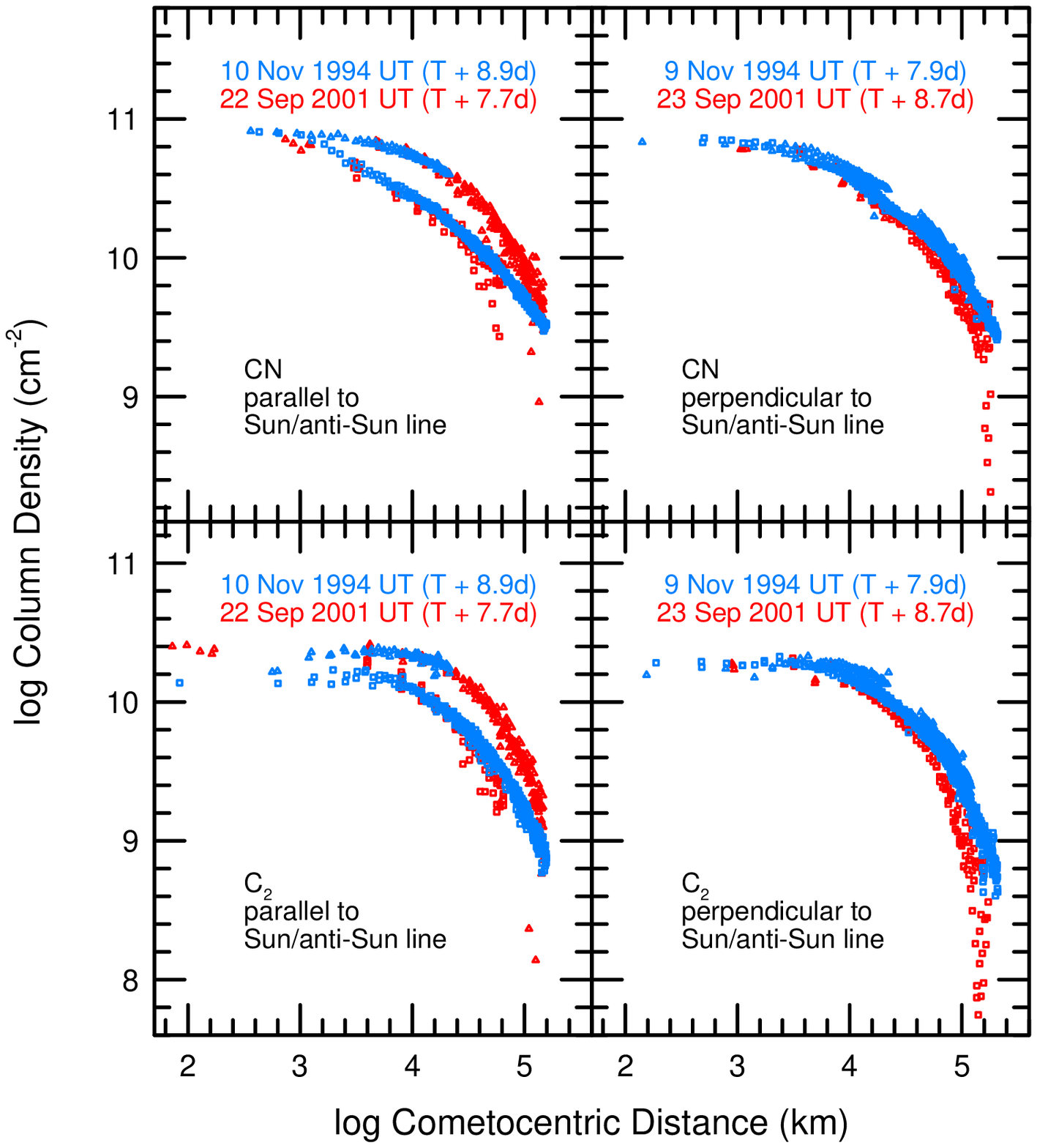}
\caption[fig5]{Farnham and Cochran}\label{compare}
\end{figure}

\noindent
{\bf Figure~\ref{compare}:} The distribution of the CN (top two
panels) and C$_{2}$ (bottom two panels) gas from the 2001 apparition
(red) compared with corresponding observations from 1994 (blue)
(Cochran and Barker 1999).  The data from the two apparitions are
plotted on the same absolute scales.  Data on the sunward side in the
lefthand panels are plotted as triangles; the tail data are plotted as
squares.  For the righthand panels, triangles and squares are used for
opposite sides of the optocenter.  The agreement between the
apparitions is quite remarkable, especially for the data obtained
parallel to the Sun/anti-Sun line.  The discrepancy in the outer coma
of the perpendicular data is discussed in the text.

\begin{figure}[p]
\vspace{7.75in}
\includegraphics{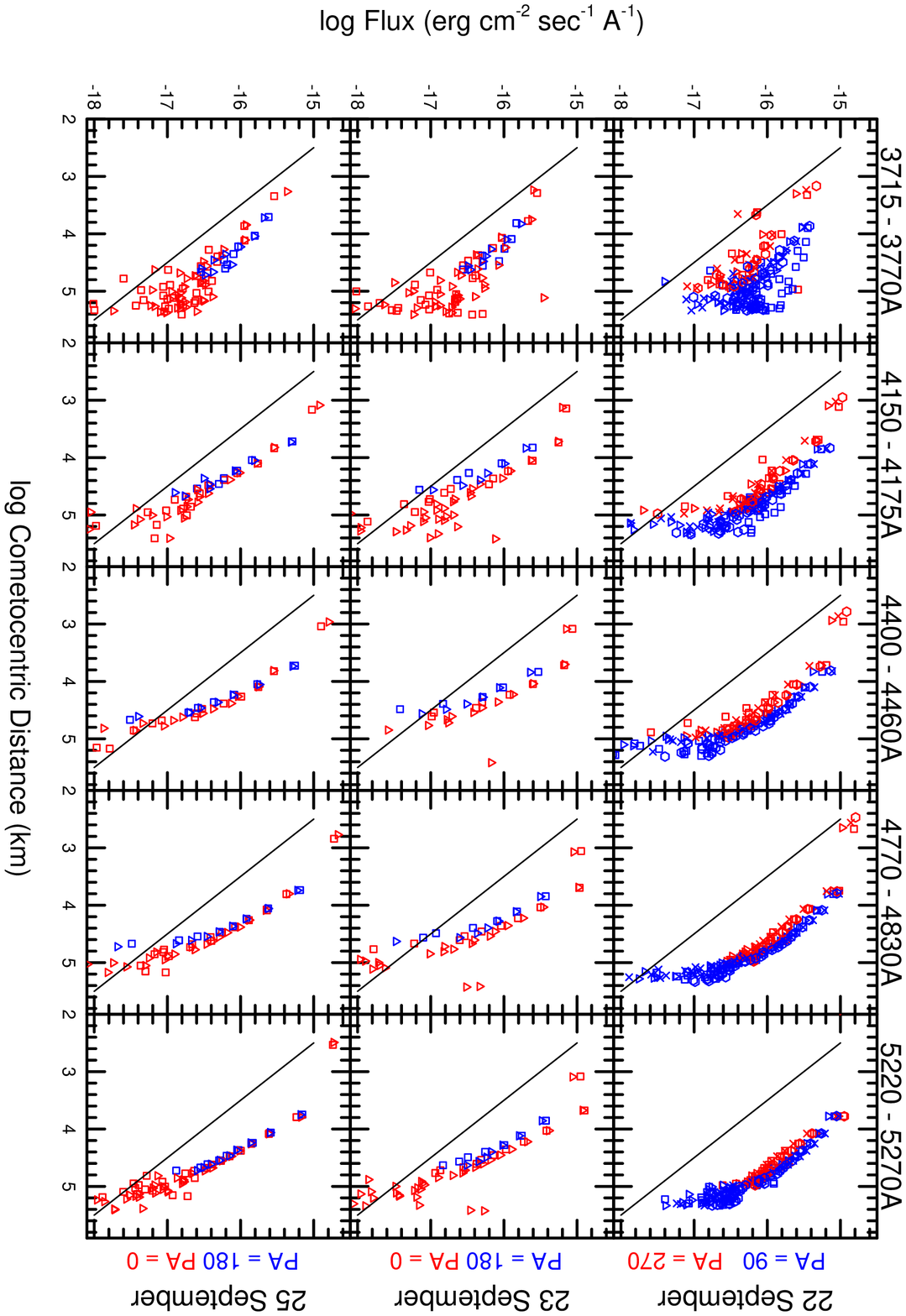}
\vspace{0.25in}
\caption[fig6]{Farnham and Cochran}\label{dustflux}
\end{figure}

\noindent
{\bf Figure~\ref{dustflux}:} The distribution of the average flux in a
continuum bandpass as a function of wavelength and cometocentric
distance.  Each horizontal row shows the continuum flux at five
bandpasses on a single night in September.  Individual spectral images
on a given night are denoted by different symbols.  The orientation of
the slit is encoded by the color with the coding at the left end of
each row.  (In the 22~September spectra, the slit was aligned
approximately along the sunward/anti-sunward direction, which is also
the direction of the primary jet.  On the other two dates, it was
perpendicular to this direction.)  The solid lines indicate a
$\rho^{-1}$ trend.  Error bars have been left off for clarity, but the
uncertainties can be estimated from the scatter in the data points.

\begin{figure}[p]
\vspace{8in}
\includegraphics{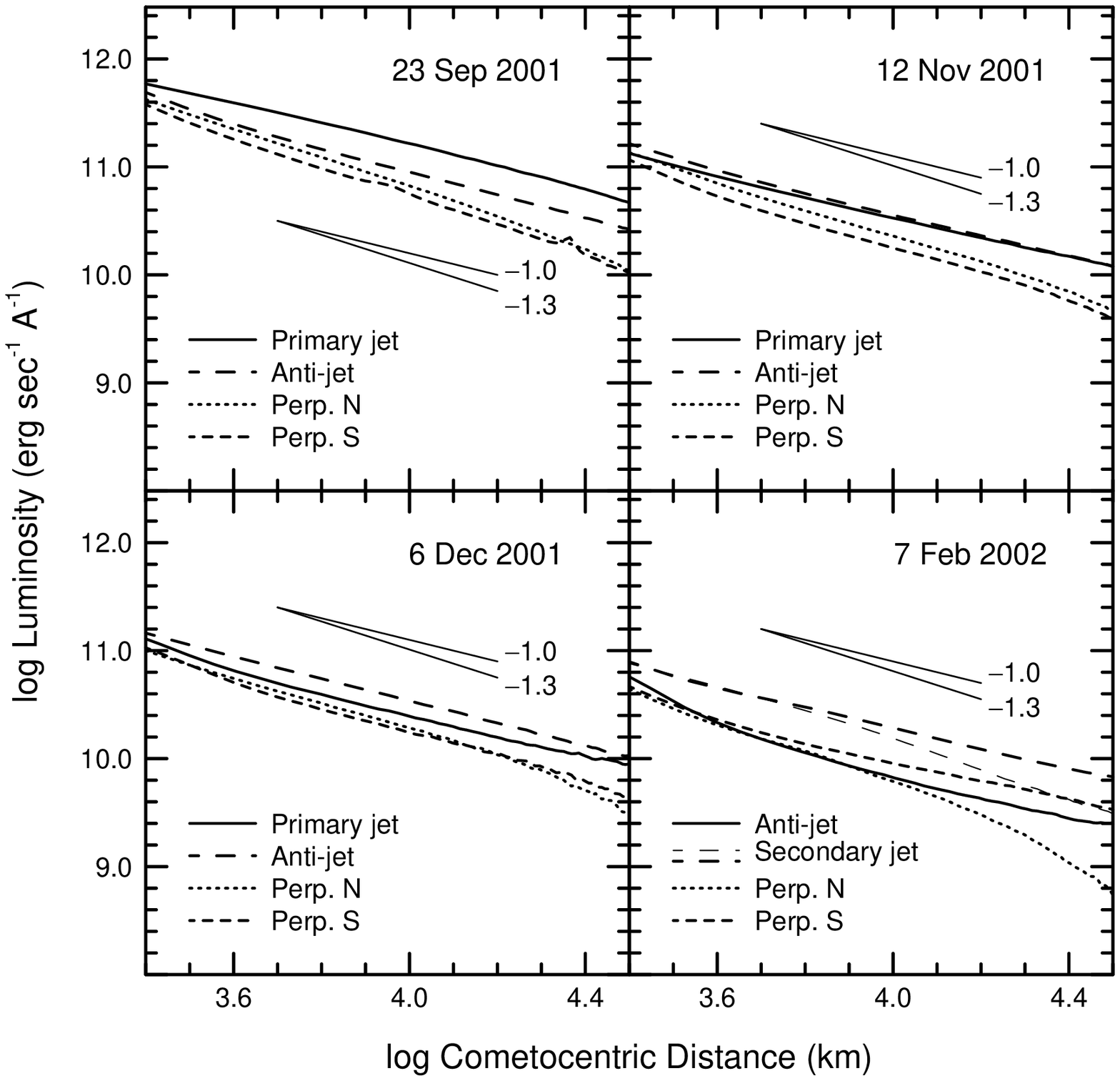}
\caption[fig7]{Farnham and Cochran}\label{radprof4}
\end{figure}

\noindent
{\bf Figure~\ref{radprof4}:} Radial profiles of the dust (extracted
from the images) on four different dates.  The four curves represent
the different profiles along the primary or secondary jet, in the
direction opposite to the jet, and in the two perpendicular
directions.  Note that the secondary jet is essentially in the
opposite direction from the primary jet; therefore the profile for the
secondary jet is plotted with the same line style as the primary
anti-jet, so the line styles are consistent with the general
direction.  In addition, there are two profiles for the secondary jet
in February.  The thin line depicts the true radial profile at a PA of
335$^\circ$, while the heavy line shows the profile following the
curvature of the jet.  All of the profiles are plotted on the same
scale, so the luminosities on different dates can be directly
compared.  Seeing variations and tracking errors affect the region
where ${\rho}<2,500$\,km in September and November, while in December
and February, when seeing was worse, the region ${\rho}<4,000$\,km is
affected. Small bumps in the profiles are caused by the profile
crossing star trails.  Slopes of -1.0 and -1.3 are denoted by the
straight lines.

\begin{figure}[p]
\vspace{8in}
\includegraphics{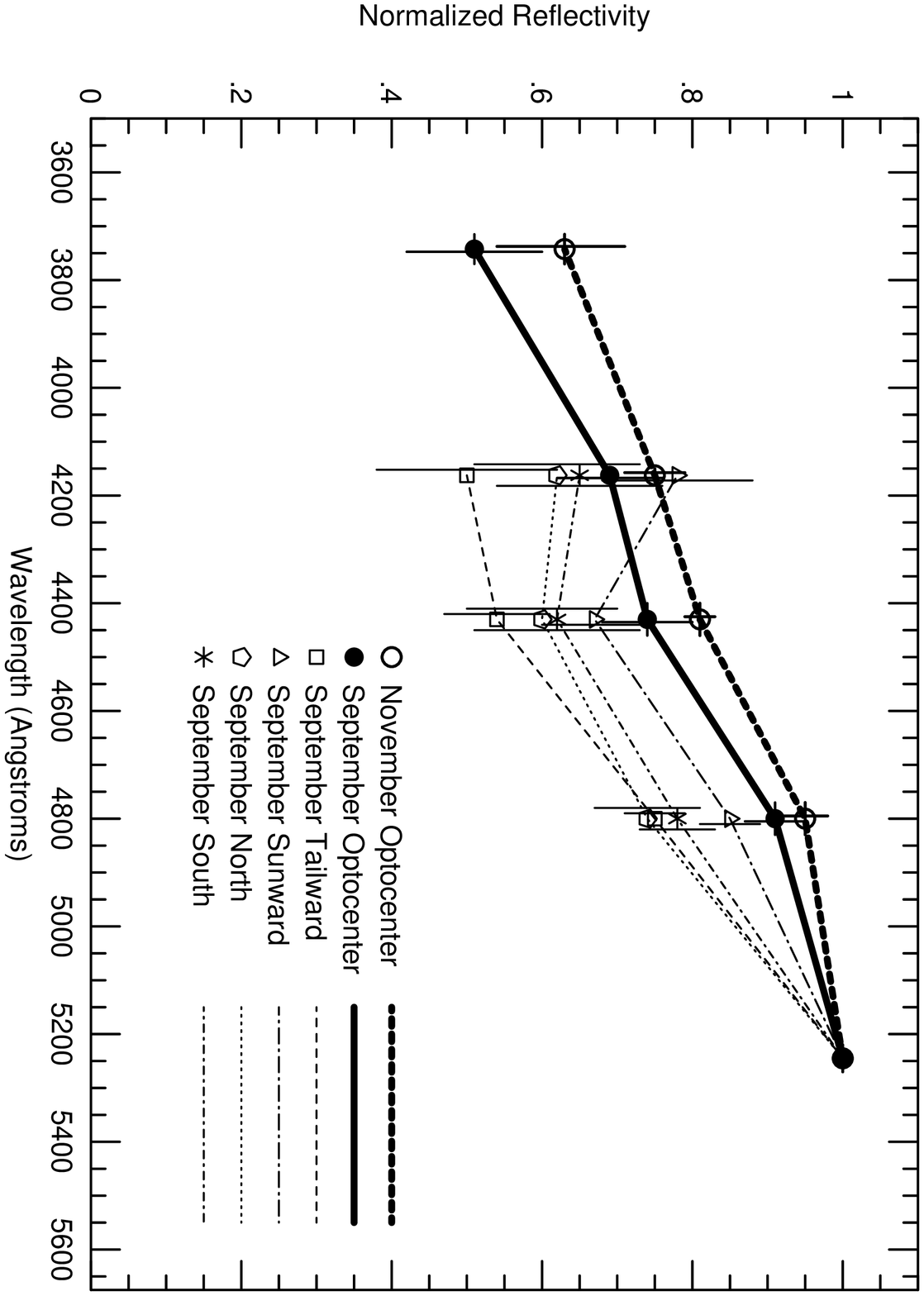}
\caption[fig8]{Farnham and Cochran}\label{reflectivity}
\end{figure}

\noindent
{\bf Figure~\ref{reflectivity}:} The reflectivities of the dust in the
coma of comet Borrelly.  This plot shows, for various positions in the
coma, the ratios of the fluxes in the continuum bandpasses of the
comet observations to those from a solar analogue star.  The
optocenter observations are the mean of the value in the optocenter
pixel for all of the observations for a given observing run.  The
directional reflectivities are the means for the three pixels just off
the optocenter in the given direction for any spectral images which
contain that orientation.  The vertical error bars are the standard
deviations from the mean and are offset to right and left of the
central wavelength for the purposes of clarity.  The horizontal bars
on the optocenter data denote the widths of the bandpasses.  All
reflectivities are normalized at 5245\AA.

\begin{figure}[p]
\vspace{8in}
\includegraphics{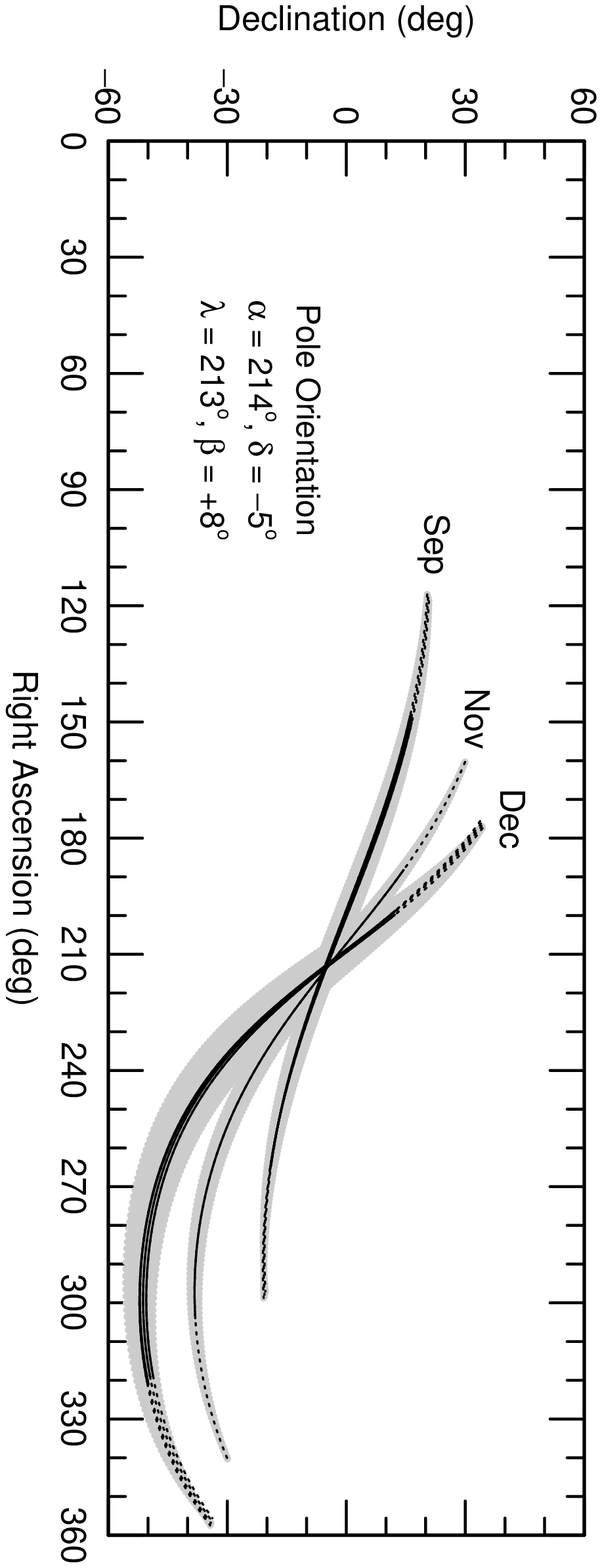}
\caption[fig9]{Farnham and Cochran}\label{poleplot}
\end{figure}

\noindent
{\bf Figure~\ref{poleplot}:} Pole/line-of-sight planes from the September,
November and December epochs as projected onto the celestial sphere.
The intersection point at $\alpha=214^\circ,~\delta=-5^\circ$ defines
the direction of the rotation axis, with an uncertainty of 4$^\circ$.
The grey outlines denote the uncertainty in the PA measurements of the
jet, and the dashed portions of each curve represent the regions of
the plane where the pole points to within 30$^\circ$ of the line of
sight.

\begin{figure}[p]
\vspace{8in}
\includegraphics{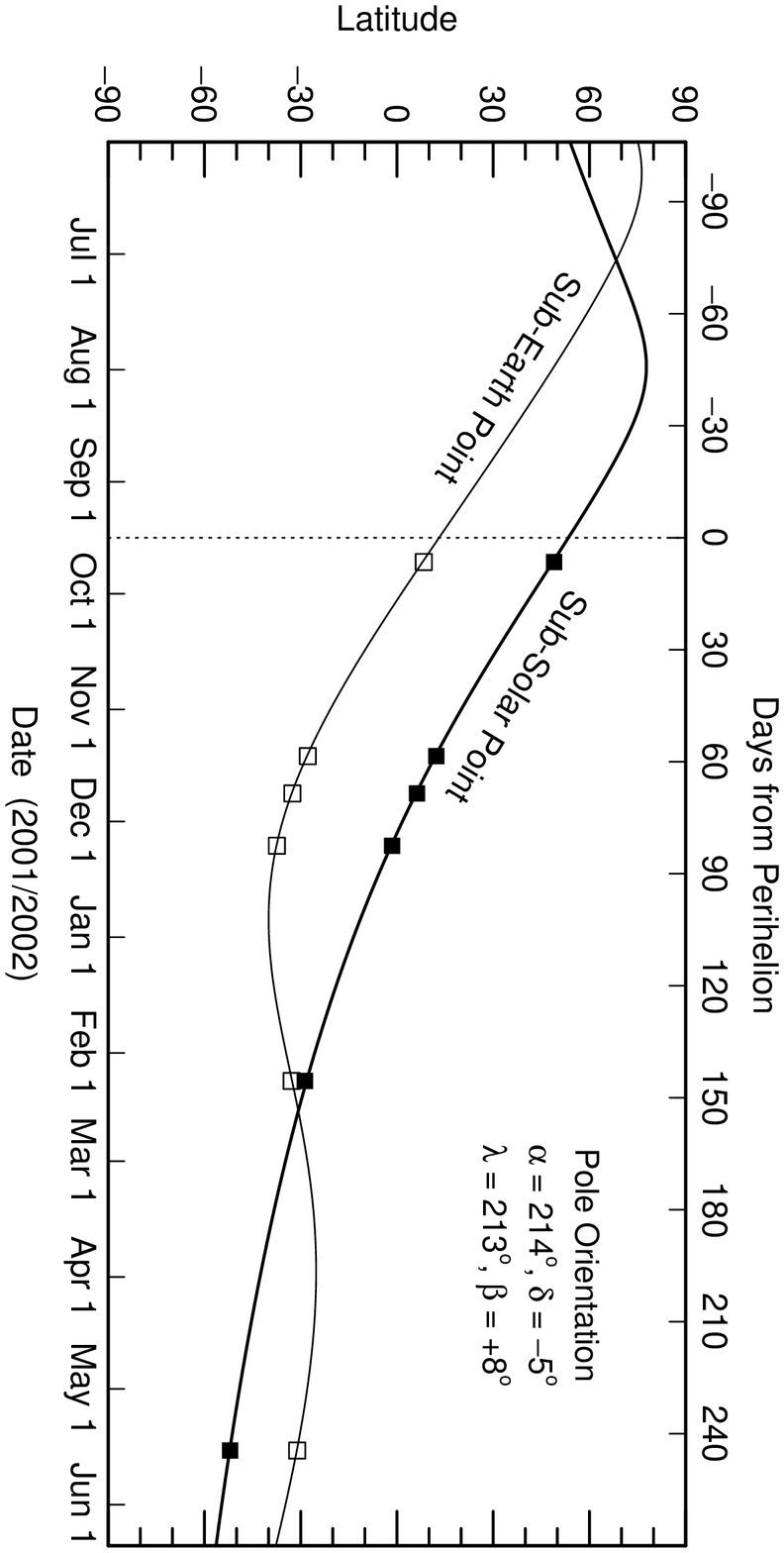}
\caption[fig10]{Farnham and Cochran}\label{subearth}
\end{figure}

\noindent
{\bf Figure~\ref{subearth}:} Plot of the comet's sub-solar and
sub-Earth latitudes for the given pole orientation as a function of
time. The northern hemisphere is defined as the one that contains the
primary jet.  The squares denote the dates on which our observations
were obtained.

\begin{figure}[p]
\vspace{8in}
\includegraphics{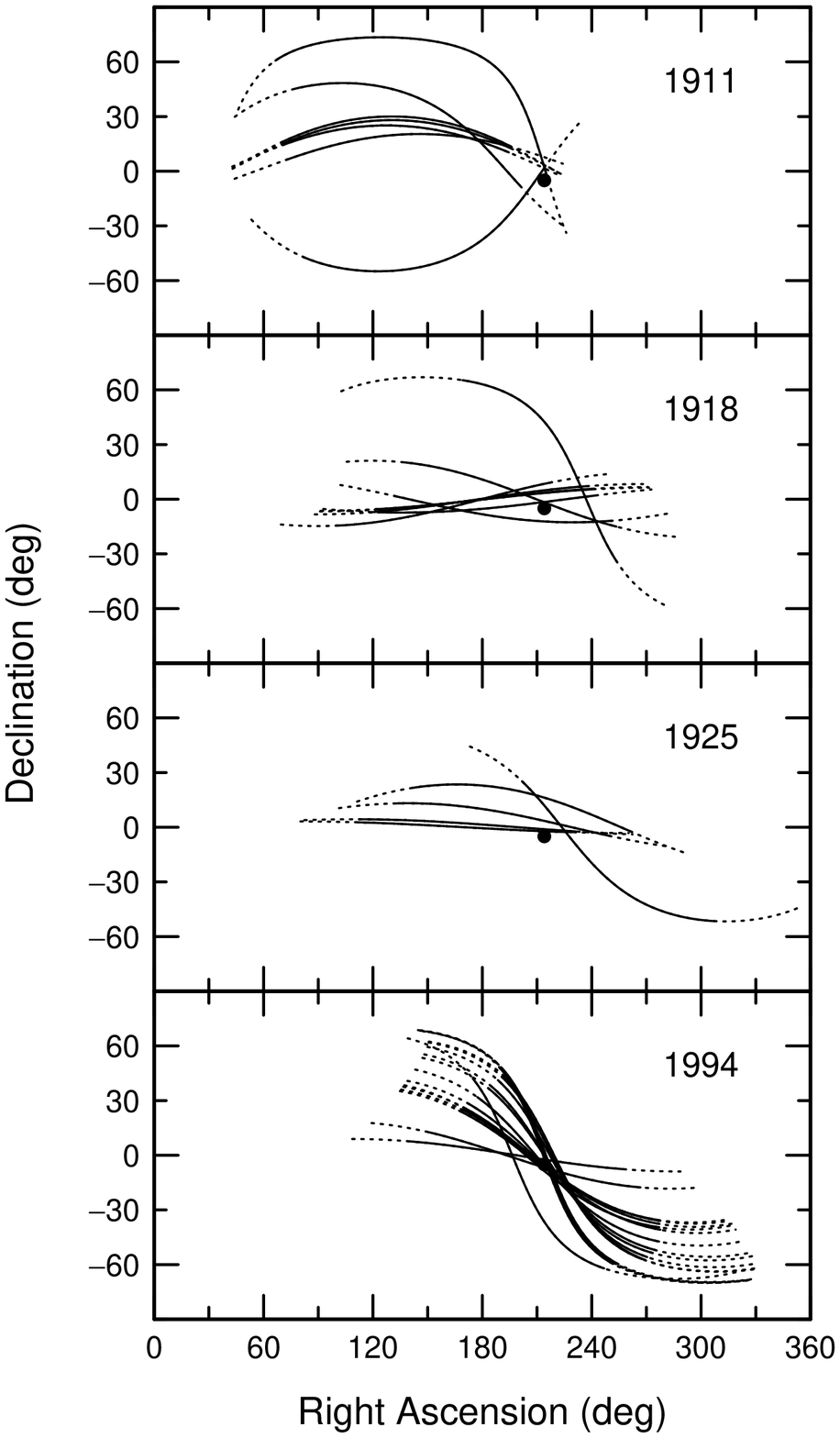}
\caption[fig11]{Farnham and Cochran}\label{sekpole}
\end{figure}

\noindent
{\bf Figure~\ref{sekpole}:} Pole/line-of-sight planes projected onto
the celestial sphere for observations obtained on previous
apparitions. The 1911, 1918 and 1925 data come from Table~IV of
Sekanina~(1979) and data from the 1994 apparition is from Fulle {\it
et al.}~(1997).  The dot in each panel represents, for comparison, the
pole position found from our 2001 data.  See the caption for
Figure~\ref{poleplot} for additional information.

\end{document}